\documentclass[a4paper,twocolumn,accepted=2021-04-12]{quantumarticle}
\pdfoutput=1

\usepackage[utf8]{inputenc}
\usepackage{amsmath,graphicx, dsfont, bm, color}
\usepackage[normalem]{ulem}
\usepackage{amssymb}
\usepackage{hyperref}

\usepackage[backend=bibtex,sorting=none]{biblatex}
\newcommand{\mean}[1]{\langle {#1} \rangle}
\newcommand{\ket}[1]{\left| {#1} \right\rangle}
\newcommand{\bra}[1]{\left\langle {#1} \right|}

\newcommand{\prjct}[1]{\mathinner{|{#1}\rangle}\!\!\mathinner{\langle{#1}|}}

\newcommand{\Id}{\mathds{1}}

\newcommand{\cD}{\mathcal{D}}

\renewcommand{\t}[1]{\mathrm{#1}}
\newcommand{\be}{\begin{equation}}
\newcommand{\ee}{\end{equation}}

\newcommand{\C}{\textrm{C}}
\newcommand{\Si}{\textrm{S}}
\newcommand{\rv}[1]{\textsf{{\itshape #1}}} 
\newcommand{\entVN}{H}
\newcommand*\colvec[1]{\begin{pmatrix}#1\end{pmatrix}}
\newcommand{\eq}[1]{Eq.~(\ref{eq:#1})}

\begin{document}

\title{Device-independent quantum key distribution from generalized CHSH inequalities}
\author{Pavel Sekatski}
\affiliation{Department of Physics, University of Basel, Klingelbergstrasse 82, 4056 Basel, Switzerland}
\author{ Jean-Daniel Bancal}
\affiliation{Department of Applied Physics, University of Geneva,
Chemin de Pinchat 22, 1211 Geneva, Switzerland}
\author{Xavier Valcarce}
\affiliation{Universit\'e Paris-Saclay, CEA, CNRS, Institut de physique th\'eorique, 91191, Gif-sur-Yvette, France}
\author{Ernest Y.-Z. Tan}
\affiliation{Institute for Theoretical Physics, ETH Zürich, 8093 Zürich, Switzerland}
\author{Renato Renner}
\affiliation{Institute for Theoretical Physics, ETH Zürich, 8093 Zürich, Switzerland}
\author{Nicolas Sangouard}
\affiliation{Department of Physics, University of Basel, Klingelbergstrasse 82, 4056 Basel, Switzerland}
\affiliation{Universit\'e Paris-Saclay, CEA, CNRS, Institut de physique th\'eorique, 91191, Gif-sur-Yvette, France}

\date{April 12, 2021}

\begin{abstract}
Device-independent quantum key distribution aims at providing security guarantees even when using largely uncharacterised devices. In the simplest scenario, these guarantees are derived from the CHSH score, which is a simple linear combination of four correlation functions. We here derive a security proof from a generalisation of the CHSH score, which effectively takes into account the individual values of two correlation functions. We show that this additional information, which is anyway available in practice, allows one to get higher key rates than with the CHSH score. We discuss the potential advantage of this technique for realistic photonic implementations of device-independent quantum key distribution.
\end{abstract}

\maketitle

\section{Introduction}

The aim of quantum key distribution (QKD) is to give two parties — Alice \& Bob — the possibility to generate a secret key when they share a quantum channel. For instance, in the implementation proposed by Ekert~\cite{Ekert91}, the channel consists of a source producing entangled particles that are distributed to Alice \& Bob. At each round, each of Alice \& Bob measure one particle by choosing one out of several measurement settings. The claim that Alice’s measurement results are secure, i.e. unknown to any third party -- Eve -- who may control the quantum channel, is guaranteed by inferring (from Alice and Bob's measurement results) that the source emits states close to pure bipartite entangled states. This ensures at the same time that Bob's results are correlated to Alice’s ones if he chooses an appropriate measurement setting, i.e. Alice and Bob’s measurement results can form a secret key. \\

Ekert suggested that the information about the key that may be available to an adversary can be quantified by choosing settings allowing Alice \& Bob to violate a Bell inequality. This idea was later progressively formalised and led to what is now called device-independent QKD (DIQKD). In its simplest version, DIQKD is implemented by letting Alice choose randomly between two measurement settings at each round, $A_x$ where $x \in \{0,1\}$, while Bob's measurement includes three possible settings, $B_y$ where $y \in \{0,1,2\}.$ For settings $x,y \in \{0,1\},$ the results -- which can possibly take many values -- are post-processed locally and turned into binary values $\rv{A}_x, \rv{B}_y \in \{-1, +1\}$. After several iterations, Alice and Bob communicate classically to estimate the \emph{CHSH score} 
\begin{equation}
S=\mean{A_0 \otimes (B_0 + B_1) + A_1 \otimes (B_0 - B_1)}
\end{equation}
where $\langle A_x \otimes B_y\rangle = p(\rv{A}_x=\rv{B}_y | x, y) - p(\rv{A}_x \neq \rv{B}_y | x, y)$ quantifies the correlation between the outcomes for measurement choices $x$ and $y$, respectively. The remaining measurement setting $y=2$ is chosen to generate an outcome $\rv{B}_2$ that minimises the uncertainty with respect to $\rv{A}_0$. Alice then forms the raw key from the outcomes $\rv{A}_0$ of the pairs that Bob measured with respect to $y=2$. \\
 
We consider $n$ such rounds, over which the source produces a tripartite state $|\Psi_{ABE}\rangle$ shared between Alice, Bob and Eve. Ref.~\cite{ArnonFriedman18} showed that Eve's information is the same as in the case where the devices have no memory and behave identically and independently in each communication round of the protocol, up to corrections vanishing with $n$. In particular, we can write $|\Psi_{ABE}\rangle = |\psi\rangle_{ABE}^{\otimes n}$ where $|\psi\rangle_{ABE}$ is the tripartite state of a single round and consider the case where measurements are done successively on the state $|\psi\rangle_{ABE}.$\\   

In the asymptotic limit of large $n$, the number of secret bits per round obtained after one-way error correction and privacy amplification (i.e.~the key rate) is then given by~\cite{DevetakWinter}
\begin{equation} 
r =  \entVN(\rv{A}_0|E) - H(\rv{A}_0|\rv{B}_2),
\label{keyrate}
\end{equation}
where $\entVN$ is the von Neumann entropy. Ref.~\cite{Acin06} showed that the conditional entropy $\entVN(\rv{A}_0|E)$ optimized over all states $\psi_{ABE}$ and measurements $A_x, B_y$ compatible with the observed CHSH score $S$ is lower bounded by 
\begin{equation}
\label{formula_pironio}
    \entVN(\rv{A}_0|E) \geq 1- h\left(\frac{1+\sqrt{(S/2)^2-1}}{2} \right)
\end{equation}
where $h$ denotes the binary entropy. This provides a lower bound on the key rate, as the conditional entropy $H(\rv{A}_0|\rv{B}_2)$ can be estimated directly from Alice and Bob measurement results associated to setting choices $A_0$ and $B_2.$ Interestingly, this bound is obtained \emph{device-independently}, i.e. without assumptions on the dimension of quantum states and the calibration of measurements. This is not the case for standard (non-device independent) QKD protocols which are not based on the violation of a Bell inequality and whose security guarantees rely on the assumption that the source and measurements carry out precisely the operations foreseen by the protocol. This assumption is hard to meet in practice and leads to vulnerabilities, as demonstrated by hacking experiments~\cite{Zhao08,Lydersen10,Weier11,Garcia19}. The robustness of device-independent quantum key distribution against these attacks makes it appealing, and a race between several experimental groups is ongoing to report the first proof-of-principle distribution of a key with a fully device-independent security. Measurement-DIQKD, a precursor of DIQKD where device-independence only applies to the measurement devices, but not to those used for state preparation~\cite{Braunstein12,Lo12}, already admits a number of experimental implementations~\cite{Rubenok13,Ferreira13,Liu13,Tang14,Comandar16}.\\

Let us note that the proof leading to the bound given in Eq.~\eqref{formula_pironio} only uses the knowledge of the CHSH score. This score is computed as a linear combination of the correlation functions $\langle A_x B_y \rangle$, but the additional information provided by considering these correlations individually -- which is anyway available in practice -- might help to facilitate a realisation of device-independent quantum key distribution. This motivation is at the core of this work. \\

Concretely, we consider the individual values of two terms appearing in the CHSH score, namely
\be
\begin{split}
    \text{X} &= \mean{A_0\otimes(B_0+B_1)}, \\
\text{Y} &= \mean{A_1\otimes (B_0-B_1)}.
\end{split}
\label{eq: X and Y}
\ee
The use of values of X and Y proved to be useful for device-independent state certification by improving the certified fidelity from the CHSH score~\cite{Valcarce2020_tobepublished}. It is also expected to be useful in DIQKD, as the knowledge of X and Y allows one to differentiate the contributions of the key generating measurement $A_0$ from the ones associated with $A_1$, from which no key is generally extracted (see~\cite{Schwonnek20} for a noticeable exception). Finally, in implementations of DIQKD with non-unit detection efficiencies where no-detection events are attributed a fixed value $\pm1$, no-detections on Bob's side can only contribute to one of these two correlation functions (X or Y). The goal of the following sections is to derive a tight bound on Eve's entropy in terms of the expected values X and Y, like the bound in Eq.~\eqref{formula_pironio} is a function of the CHSH score $S$. The main result of this work is to confirm the intuition that the use of individual values of X and Y improves the bounds on Eve's information derived from the CHSH score, and hence the key rate of DIQKD. We also apply the new bound to a concrete setup using a photon pair source based on spontaneous parametric down conversion (SPDC) and photon detections and show that it leads to a subtantial improvement of the key rate, at least for high detection efficiency.\\


\section{Formulation of the problem}
\paragraph{Generalization of the CHSH test--} We are interested in bounding Eve's conditional entropy $H(\rv{A}_0|E)$ appearing in Eq.~\eqref{keyrate} as a direct function of observed quantities X and Y. Formally, this can be accomplished by considering all possible quantum models ($\psi_{ABE}, A_x, B_y $) that satisfy  $\text{X}_{\text{model}}\geq \text{X}$ and $\text{Y}_\text{model}\geq \text{Y}$. However, it is clear that the set of points (X,Y) for which Eve's conditional entropy is bounded from below by some constant is convex: two quantum models giving some amount of information to Eve can be joined into a new model on which Eve's conditional entropy is bounded by the weighted sum of entropy bounds associated to the individual models. It is thus equivalent to bound Eve's conditional entropy with linear constraints of the form
\be
\frac{ \cos(\Omega)}{2}\text{X}_\text{model} + \frac{\sin(\Omega)}{2} \text{Y}_\text{model} \geq \beta,
\ee
where $\beta$ is deduced from the observed quantities X and Y from the following formula 
\be\label{eq:bellOmega}
\beta = \frac{1}{2}\left(\cos \left(\Omega\right)\text{X}+\sin\left(\Omega\right)\text{Y}\right).
\ee
(Further in the text, we will use a compact notation for sine $\Si_\Omega=\sin(\Omega)$ and cosine $\C_\Omega = \cos(\Omega)$.) Obviously, $\Omega=\pi/4$ reduces back to the CHSH constraint (up to normalization). Just like the CHSH score can be seen as the result of a test of the CHSH inequality, we can associate $\beta$ to the test of a Bell inequality -- a generalisation of the CHSH inequality -- that is characterized in Ref.~\cite{Acin2012}.
This characterization is also done below for the sake of completeness.\\

\paragraph{Reduction to qubits--} The score $\beta$ given in Eq.~\eqref{eq:bellOmega} is estimated when Alice and Bob choose the measurement $A_x$, $B_y$, $x, y=0,1$, which are observables with eigenvalues $\pm 1.$ 
Jordan's lemma~\cite{jordan1875,Supic19} tells us that such observables can be jointly block diagonalised with blocks of size $2 \times 2$, i.e. 
\be
A_x = \bigoplus_k A_{x,k} \qquad B_y = \bigoplus_{k'} B_{y,k'}\;,
\ee
where, without loss of generality, we can assume the restriction to each qubit block to be a real Pauli measurement satisfying $A_{x,k}^2=\Id_k$ and $B_{y,k'}^2=\Id_{k'}$. This means that in each block labelled by $k$ and $k'$ respectively, the measurement is characterized by unit vectors ${\bf a}_x^k$, ${\bf b}_y^{k'}$ such that
\be
    A_{x,k} = {\bf a}_x^k \cdot \binom{\sigma_z}{\sigma_x} \qquad B_{y,k'} = {\bf b}_y^{k'} \cdot \binom{\sigma_z}{\sigma_x} \quad 
\ee
where $\sigma_z$ and $\sigma_x$ are Pauli operators.

The state $\ket{\psi}_{ABE}$ can be enforced to take the form 
\be
\ket{\psi}_{ABE} =\bigoplus_{k,k'} \sqrt{p_{(k,k')}} \ket{\psi}_{ABE}^{(k,k')},
\ee
where $p_{(k,k')}$ is a probability distribution and $\ket{\Psi}_{ABE}^{(k,k')} \in \mathds{C}^2_A \otimes \mathds{C}^2_B\otimes \mathcal{H}^{(4)}_E,$ see Refs~\cite{Pironio09,Ho20} for detailed discussions. Given models with such measurements and state, the quantity of interest can be expressed as
\be
\label{eq:convex sum}
\entVN(\rv{A}_0|E) = \sum_{k,k'} p_{(k,k')} \entVN_{(k,k')}(\rv{A}_0|E)
\ee
where $\entVN_{(k,k')}(\rv{A}_0|E)$ is Eve's conditional entropy for four-qubit models (including the two qubits from Eve's purification) involving real Pauli measurements. If the minimization of $\entVN_{(k',k)}(\rv{A}_0|E)$ over such models satisfying $\frac{\cos(\Omega)}{2}\text{X}_\text{model, k, k'} + \frac{\sin(\Omega)}{2} \text{Y}_\text{model, k, k'} \geq \beta$ provides a convex function of $\beta,$ this function can be used directly as a lower bound on the quantity $\entVN(\rv{A}_0|E)$ through Eq.~\eqref{eq:convex sum}. If it is not convex, it can be convexified so as to apply to all possible mixtures of state and measurement, and thus again apply to Eq.~\eqref{eq:convex sum}. This convexity property allows us to reduce the general problem of finding the minimum of Eve's conditional entropy over all possible models to a minimization over four-qubit models with real Pauli measurements. We will come back to this convexification requirement later.\\

\paragraph{Noisy preprocessing--} We consider a simple post-processing of the raw key, known as \emph{noisy pre-processing}~\cite{Renner05a, Renner05b, Renes07}, which has been shown to be beneficial to reduce the requirement on the detection efficiency in photonic implementations of device-independent quantum key distribution~\cite{Ho20}. Once the raw key is obtained, Alice is instructed to generate a new raw key $\widehat{\rv{A}}_0$ by flipping each bit of the initial raw key with a probability $p.$ 
(This can be described using a POVM that is a mixture of the original measurement and a measurement with the outcome labels flipped.)
Note that we will often parametrise the amount of noise that Alice adds with a parameter $q=(1-2p)^2$.\\

\paragraph{Symmetrization--} In order to simplify the analysis, it is convenient to consider a symmetrization step in which both parties, Alice and Bob, flip the outcomes of the key generating measurements depending on a public random bit string. This guarantees that bits of the raw key are random, i.e. $H(\rv{A}_0)=H(\widehat{\rv{A}}_0)=1$. Importantly, one can show the equivalence of protocols with and without symmetrization, meaning that the symmetrization does not need to be implemented in practice, see~\cite{Ho20} for a complete description of the symmetrization step in the presence of noisy preprocessing.\\

\paragraph{Reduction to Bell diagonal states--} If the constraints appearing in the minimization problem do not depend on the marginal probabilities $p(\rv{A}_x|x)$ and $p(\rv{B}_y|y)$ of Alice and Bob respectively, the symmetrization step previously presented reduces the model of the state to a Bell-diagonal structure
\be
\label{twoqubitmaxent_decomposition}
\ket{\psi}_{ABE} = \sum_{i=1}^4 \sqrt{L_i} \ket{\Phi^i}_{AB}\ket{i}_E
\ee
where $\ket{\Phi_i}=\{ \Phi^+, \Psi^-,\Phi^-,\Psi^+\}_{i=1}^4$, and without loss of generality, a partial ordering of the eigenvalues $L_1\geq L_2$ and $L_3\geq L_4$ can be imposed~\cite{Pironio09}. Note that the superscripts $k, k'$ are omitted in the tripartite state appearing in Eq.~\eqref{twoqubitmaxent_decomposition}, i.e. $\ket{\psi}_{ABE} \rightarrow \ket{\psi}_{ABE}^{(k,k')}.$ Until the end of this section and in the next section which is dedicated to the resolution of the optimization presented in Eq. \eqref{eq: direct qubit}, we remove the index $k, k'$ for making the notation simpler and ask the reader to keep in mind that we consider the restriction to four qubit models with real Pauli measurements in this two sections. 
\\

\paragraph{Eve's conditional entropy--}
Eve's conditional entropy can be expressed as
\be
\label{eq: entropy jingle}
\small
    H(\widehat{\rv{A}}_0|E) = H(\widehat{\rv{A}}_0) - H(\rho_E) + \sum_{\hat{a}=\pm 1} p(\hat{a}) H(\rho_{E|\hat{a}})
\ee
where $\rho_E$ is the reduced state of Eve and $\hat{\rho}_{E|\hat{a}}$ corresponds to Eve's state conditioned on Alice's noisy key bit $\widehat{\rv{A}}_0$ being equal to $\hat{a}$, which occurs with probability $p(\hat{a})$. The equivalence of the protocol with the symmetrized one allows us to take $H(\widehat{\rv{A}}_0)=1$ and  $p(\hat{a})=\frac{1}{2}$.

$H(\rho_E)$ is given by the entropy $H({\bf L})$ of the probability vector ${\bf L}=(L_1,\hdots,L_4)$, while for the $H(\rho_{E|\hat{a}})$ terms we have
\begin{align}
&\rho_{E|\hat{a}=+1} = \\
&{\small \left(\arraycolsep=2pt
\begin{array}{cccc}
 L_1 & 0 & \C_\phi \sqrt{L_1 L_3 q} & \Si_\phi \sqrt{L_1 L_4 q} \\
 0 & L_2 & \Si_\phi \sqrt{L_2 L_3 q} & -\C_\phi \sqrt{L_2 L_4 q} \\
 \C_\phi \sqrt{L_1 L_3 q} & \Si_\phi \sqrt{L_2 L_3 q} & L_3 & 0 \\
 \Si_\phi \sqrt{L_1 L_4 q} & -\C_\phi \sqrt{L_2 L_4 q} & 0 & L_4 \\
\end{array}\nonumber
\right),}
\end{align}
where $\phi$ labels Alice measurement $A_0 = \cos(\phi) \sigma_z + \sin(\phi) \sigma_x$ (we use the notation $\C_\phi = \cos(\phi)$ and $\Si_\phi=\sin(\phi)$). The two states $\rho_{E|\hat{a}=\pm 1}$ are related by a simple unitary transformation and therefore have the same entropy, see App.\ref{app: entropy formulas} for details. The expressions of these entropic quantities provide an explicit way to compute $H(\widehat{\rv{A}}_0|E)$ as a function of the parameters $\bf L$ and $\phi.$ Let us now turn our attention to the constraints.\\
\paragraph{Quantum correlations in the (X,Y) plane --} As mentioned earlier, we are considering quantum models with the values of correlators X and Y given by Eq.~\eqref{eq: X and Y}. Without loss of generality, we can assume $\text{X},\text{Y}\geq 0$, which can always be attained by relabelling the measurement outcomes of $A_1,B_0$ and $B_1$ (i.e. without touching the angle $\phi$).

In this positive quadrant of the plane, the local strategies are delimited by the CHSH inequality $\text{X}+\text{Y}\leq 2$, i.e. the line connecting the deterministic strategies $(\text{X},\text{Y})=(2,0)$ and $(\text{X},\text{Y})=(0,2)$. This implies the following local bounds for the generalized CHSH tests
\be
\label{generalizedBellinequality}
\frac{1}{2}\C_\Omega \text{X} + \frac{1}{2}\Si_\Omega  \text{Y}\leq B_\Omega^L = \max (\C_\Omega ,\Si_\Omega ).
\ee

To identify the upper limit of the quantum set, we  consider the expected values of the generalized CHSH operator
\be\label{eq: bell tests}
\mathcal{B}_\Omega = \left\langle \frac{\C_\Omega}{2}  A_0\otimes(B_0+B_1) + \frac{\Si_\Omega}{2} A_1\otimes (B_0-B_1) \right\rangle.
\ee
To find its maximum value, we use the qubit parametrization of measurements $A_y, B_y$ and parametrize the measurement angles on Bob's side as
\begin{eqnarray}\label{eq: B0B1 param}
    B_{0}+B_{1} &=& ({\bf b}_0+{\bf b}_1) \cdot \binom{\sigma_z}{\sigma_x} = 2 \C_\theta\,  {\bf c} \cdot \binom{\sigma_z}{\sigma_x}\ \ \ \\
    B_{0}-B_{1} &=& ({\bf b}_0-{\bf b}_1)\cdot \binom{\sigma_z}{\sigma_x} = 2 \Si_\theta\,  {\bf c}_\perp \cdot \binom{\sigma_z}{\sigma_x}\ \ \  
\end{eqnarray}
with two arbitrary perpendicular unit vectors $\bf c$ and ${\bf c}_\perp$, and $\cos(2\theta) = {\bf b}_0 \cdot {\bf b}_1$. From the diagonalization of the operator on the right hand side of Eq.~\eqref{eq: bell tests}, one finds the quantum bound
\be
\mathcal{B}_\Omega^Q=1
\ee
attained at $(\text{X},\text{Y}) = (2 \cos(\Omega), 2\sin(\Omega))$ by a maximally entangled two qubit state and measurement settings ${\bf a}_0 \cdot {\bf a}_1 =0$ and ${\bf b}_0 \cdot {\bf b}_1=\cos(2\Omega)$. It follows that Eve's information is constrained by the part of the quantum set lying between the line $\text{X}+\text{Y}=2$ and the circle $\text{X}^2+\text{Y}^2= 4$. At this point, we can already conclude that any quantum model with (X,Y) lying on the circle satisfies $H(\hat{\rv{A}}_0|E)=1$ (except for the two points with X+Y$=2$), since the underlying state of Alice and Bob has to be pure. This is a straightforward improvement over the CHSH bound.\\

\paragraph{Formulation of the problem to solve--}
The reductions introduced so far invite us to first solve the following optimization  
\be\label{eq: direct qubit}
	\small
\begin{split}
	I(\beta; \Omega,q ) = \max_{\bf{L}, \phi ,{\bf a}_1, {\bf b}_0, {\bf b}_1 } &\,     H({\bf L}) - H(\rho_{E|\hat{a}=+1})\\
\text{s.t.}\ \ \ \ &\, \mathcal{B}_\Omega({\bf L}, \phi, {\bf a}_1, {\bf b}_0, {\bf b}_1) \geq \beta
\end{split}
\ee
and then consider directly the solution $I(\beta; \Omega,q)$ if it is concave in $\beta$ or construct a concave function $\widehat{I}(\beta; \Omega,q)\geq I(\beta; \Omega,q)$ to bound Eve's uncertainty using
\be \label{eq: best HE}
H(\widehat{\rv{A}}_0|E) \geq 1- \min_\Omega \widehat{I}\left(\frac{\C_\Omega\text{X}+\Si_\Omega \text{Y}}{2};\Omega, q\right).
\ee
Note that from the symmetries of the goal function $H({\bf L}) - H(\rho_{E|\hat{a}=+1})$ and the constraint $\mathcal{B}_\Omega,$ we can assume $\phi \in[0,\frac{\pi}{4}]$ for the key generating setting and $L_1-L_2\geq L_3-L_4$ in addition to $L_1\geq L_2$ and $L_3\geq L_4$ for the state, see App.\ref{app: parametrization} for the details. Further note that we will often use a parametrisation of the tripartite state given by the following 3 component vector
\be
\left(\begin{array}{c} 
T_z \\
T_x\\
T_p
\end{array}\right)
=
\left( \begin{array}{c} 
L_1-L_2+L_3-L_4\\
L_1 -L_2 -L_3+L_4  \\
L_1 +L_2-L_3-L_4
 \end{array}\right),
\ee
with $0\leq T_x\leq T_z\leq 1$ and $ T_z+T_x- 1\leq T_p \leq 1- (T_z-T_x)$.


\section{Bounding Eve's information with generalized CHSH tests}

We are now ready to compute a bound on Eve's information as a function of the generalized CHSH score given in Eq.~\eqref{eq:bellOmega} by solving the optimisation problem given in Eq.~\eqref{eq: direct qubit}. Among the parameters of the model in Eq.~\eqref{eq: direct qubit}, the measurement setting ${\bf a}_1, {\bf b}_0$ and ${\bf b}_1$ only influence the constraint but not the goal function. Furthermore, it is shown in Ref.~\cite{Ho20} that $H(\rho_{E|\hat{a}=+1})$ is a monotonic function in the key generating setting $\phi \in [0,\frac{\pi}{4}]$. We can thus decompose the maximization problem in two steps. First, for a fixed state $\bf L,$ we find the lowest angle $\phi$ 
allowing to satisfy the constraint 
\be\label{eq: best angle}\begin{aligned}
	\phi_*({\bf L},\beta, \Omega) = \min&\ \ \phi \\
    \text{s.t.}& \max_{{\bf a}_1, {\bf b}_0, {\bf b}_1 } \mathcal{B}_\Omega({\bf L}, \phi, {\bf a}_1, {\bf b}_0, {\bf b}_1) \geq \beta.
\end{aligned}
\ee
Second, we fix $\phi = \phi_*({\bf L},\beta, \Omega)$ to the optimal value for Eve, and maximize her information with respect to the state, that is, we solve
\be
\small
	I(\beta; \Omega,q ) = \max_{\bf{L}} H({\bf L})
	- H(\rho_{E|\hat{a}=+1}; \phi_*({\bf L},\beta, \Omega)).
\ee

We solve Eq.~\eqref{eq: best angle} in App.~\ref{app: best phi}. The expression of the optimal angle $\phi_*$  depends on whether the parameter $\Omega$ exceeds $\frac{\pi}{4}$.
We treat the two cases $\Omega\leq\frac{\pi}{4}$ and $\Omega>\frac{\pi}{4}$ separately.\\

\subsection{The simple case with $\Omega\leq\frac{\pi}{4}$}
\label{sec: simple case}

For the Bell tests satisfying $\Omega\leq\frac{\pi}{4}$, which include the CHSH test, the observed score $\beta$ does not constrain the key generating setting $\phi$ but only the state $\bf L$, see App.~\ref{app: best phi} for details. As a result, there always exists a realization with the optimal angle $\phi_*=0$ as long as the state is such that the Bell score can be attained, i.e. if
\be
\C_\Omega^2 T_z^2 +\Si_\Omega^2 T_x^2 \geq  \beta_\Omega^2.
\ee

In other words, the optimization Eq.~\eqref{eq: best angle} yields $\phi_*=0$, and the maximization of the entropy becomes possible: the conditional state $\rho_{E|\hat{a}=+1}$ is then block diagonal and its entropy has a simple closed-form expression.
Such a maximization has been done for the CHSH case ($\Omega=\frac{\pi}{4}$) in Ref.~\cite{Ho20}, but Ref.~\cite{Woodhead20} pointed out that the analytical bounds on conditional entropies given in this reference assume qubit attacks. The same bounds were derived using a different approach in~\cite{Woodhead20}, where it is also proved that these bounds are convex. The convexity results of~\cite{Woodhead20}, together with Jordan’s lemma, imply that the obtained qubit bounds are in fact valid for any dimensions. The same convexity proof applies to the current situation with $\Omega\leq\pi/4$. For the sake of completeness, we provide in App.~\ref{app: convex} an alternative proof of convexity, which directly applies to the present case and to~\cite{Ho20}. We show in particular (see App.~\ref{app: analytic entropy}) that

\be
\label{easy_sol}
\begin{split}
I(\beta; \Omega,q ) = h_q(z) & = h(z) - h(n_q(z)) \\
 \text{with} \quad n_q(z) &=\frac{1 + \sqrt{1 - 4\, (1-q)\, z (1-z)}}{2}\\
 \text{and\ \ \ \ \ \,} \quad z&= \frac{1}{2} \left(\frac{\sqrt{ \beta ^2-\C_{\Omega}^2}}{ \Si_\Omega }+1\right),
\end{split}\ee
where $h$ is the binary entropy. The concavity of $I(\beta; \Omega,q )$ and hence the convexity of Eve's entropy $H(\widehat{\rv{A}}_0|E)$ follows from (see App.~\ref{app: convex})
\be\label{eq: convex proof}
\frac{d^2}{d\beta^2} h_q\big(z(\beta)\big) = h_q''(z) \big(z'(\beta)\big)^2 + h_q'(z)z''(\beta)\leq 0.
\ee

Finally, it remains to determine the optimal inequality to use for a given point (X,Y).  $h_q(z)$ being a monotonic function of $z$, we want to maximize its argument
\begin{eqnarray}
z &=& \frac{1}{2} \left(\frac{\sqrt{ (\frac{\C_\Omega \t{X} + \Si_\Omega \t{Y}}{2})^2-\C_{\Omega}^2}}{  \Si_\Omega }+1\right)\\
&=& \frac{1}{4}\sqrt{\frac{4 \,\t{Y}^2}{4-\t{X}^2}- (4-\t{X}^2)\left(\cot(\Omega)-\frac{\t{X Y}}{4-\t{X}^2}\right)^2}+\frac{1}{2},\nonumber
\end{eqnarray}
with respect to $\Omega$ in the range where the argument of the square root is positive (which means that the value for the Bell test exceeds the local bound). Manifestly, the expression has a global maximum at
\be\label{eq:bestOmega}
\cot(\Omega) = \frac{\t{X} \t{Y}}{4-\t{X}^2}.
\ee
Now, we have to verify that the optimal test we found satisfies $\Omega\leq \frac{\pi}{4}$. One finds that this is the case for
\be
\frac{4-\text{X}^2}{\text{XY}} \leq 1, 
\ee
providing a bound on Eve's entropy as a direct function of the correlators X and Y:
\be
\quad z_\text{opt}=  \frac{1}{2} \left(\frac{\text{Y}}{\sqrt{4-\text{X}^2}}+1\right).
\ee
One easily verifies that this bound is indeed better than the CHSH formula~\cite{Ho20}
\be
\quad z_\text{CHSH}=  \frac{1}{2} \left(1+\sqrt{\left(\frac{\text{X}+\text{Y}}{2}\right)^2-1}\right)
\ee
if $\frac{4-\text{X}^2}{\text{XY}}<1$.


\subsection{The complicated case with $\Omega >\frac{\pi}{4}$}
\label{sec: complicated}
\subsubsection{Bounding Eve's information from a numerical optimization}
For the remaining Bell tests, with $\Omega >\frac{\pi}{4}$, the situation is different. Here, the generalized CHSH score $\beta$ does not only constraint the state $\bf L$ but also the setting of the key generating measurement. We therefore adopt a strategy in two steps. First, we develop a method that can efficiently compute a bound on Eve's information, either heuristically or under a set of well-defined ansatz. Second, we provide a numerical method able to certify formally the validity of a given bound.

Considering the parameters when $\Omega >\frac{\pi}{4}$, we find that there are two different regions. First, for the states falling in the range
\be
\mathds{S}(\beta,\Omega)=\{{\bf L}| (\C_\Omega^2 T_z^2 +\Si_\Omega^2 T_x^2) \leq \beta^2 \leq (\C_\Omega^2 T_x^2 +\Si_\Omega^2 T_z^2)\},
\ee
the constraint $\mathcal{B}_\Omega\geq \beta$ can be satisfied with the measurement angle $\phi\geq \phi_*({\bf L},\Omega,\beta)$, where $\cos^2(\phi_*) = c_*^2({\bf L},\Omega,\beta)$ is given by
\be
c_*^2({\bf L},\Omega,\beta)=\frac{(\beta^2 -\Si_\Omega^2 T_x^2)(\C_\Omega^2 T_x^2 +\Si_\Omega^2 T_z^2 -\beta^2)}{\C_\Omega^2(T_z^2-T_x^2)(\Si_\Omega^2 T_z^2 +\Si_\Omega^2 T_x^2 -\beta^2)}.
\ee
The constraint on the angle only becomes trivial, $c_*^2=1$, on the boundary of the region $\mathds{S}$, where $(\C_\Omega^2 T_z^2 +\Si_\Omega^2 T_x^2) = \beta^2$. 

Second, in the region where $(\C_\Omega^2 T_z^2 +\Si_\Omega^2 T_x^2) > \beta^2$  the Bell score $\beta$ can also be attained with $\phi=0$. However, there models provide less information to Eve as compared to these on the boundary $(\C_\Omega^2 T_z^2 +\Si_\Omega^2 T_x^2) = \beta^2$, see the discussion at the end of App.~\ref{app: analytic entropy}. So we can safely ignore this region.

To find the best strategy for Eve, it thus remains to solve
\be\label{eq: numerical solution}
I(\beta;\Omega,q) = \max_{{\bf L}\in \mathds{S}(\Omega,\beta)} H({\bf L}) - H(E|\hat{a}=+1;c_*^2({\bf L},\Omega,\beta)).
\ee
This optimization only involves an analytic function of three parameters on a compact domain. It can be easily and time-efficiently solved heuristically by standard numerical methods, e.g using fmincon on MATLAB, NMaximize on Mathematica, or scipy.optimize on Python. We now give an ansatz on the solution of the optimization given in Eq.~\eqref{eq: numerical solution}, which allows one to speed up its numerical resolution even further.\\

\subsubsection{Ansatz}
\label{sec:anzatz}

First, we observe that the vector ${\bf L}$ saturating Eve's information only has two non-zero coefficients $L_1=1-L_3$ and $L_2=L_4=0$, or $T_z=1$ and $T_x=T_p$. With this observation, the previous optimization problem becomes a scalar optimization, that is 
\be
\begin{split}
    \tilde I_\text{anz}(\beta;\Omega,q) = &\max_{ \frac{\beta^2 -\Si_\Omega^2}{\C_\Omega^2}\leq T_x^2\leq  \frac{\beta^2 -\C_\Omega^2}{\Si_\Omega^2}}\,  h\left(\frac{1+T_x}{2}\right) \\
    &- h\left(\frac{1+\sqrt{T_x^2+c_*^2q (1-T_x^2)}}{2}\right).
\end{split}
\ee

Second, we see that the bound $\tilde I_\text{anz}(\beta;\Omega,q)$ is not concave for small $\beta$. However, we observe that its concave roof can be obtained by drawing a line which passes through the point 
\be\begin{split}
\beta&=\mathcal{B}_\Omega^L= \sin(\Omega)\\
I_L(q)&=I(\mathcal{B}_\Omega^L; \Omega, q)=1-h\left(\frac{1+\sqrt{q}}{2}\right)
\end{split}\ee
which is tangent to the curve $\tilde  I_\text{anz}(\beta;\Omega,q)$. The value of the generalized CHSH score at the tangent point can be found by solving
\be\label{eq: beta*}
\beta_* = \underset{\beta}{\text{argmin}} \frac{I_L(q)-\tilde  I_\text{anz}(\beta;\Omega,q)}{\beta-\sin(\Omega)}.
\ee
Labeling $I_*(\Omega,q)= \tilde I_\text{anz}(\beta_*;\Omega,q)$, this leads to the concave roof
\be\label{eq: I tilde}
\widehat{I}_\text{anz}(\beta;\Omega,q) =\begin{cases}\frac{I_*(\Omega,q) (\beta-\sin(\Omega))-  I_L(q)(\beta-\beta_*)}{\beta_*-\sin(\Omega)} & \beta < \beta_*\\
\tilde I_\text{anz}(\beta;\Omega,q) & \beta \geq \beta_*.
\end{cases}
\ee

At this stage, we further observe that the optimal value of $T_x^2$ for values of $\beta\geq \beta_*$ coincides with its maximum possible value $T_x^2=\frac{\beta^2 -\C_\Omega^2}{\Si_\Omega^2}$ (implying $c_*=1$). We thus define $I_\text{anz}(\beta;\Omega,q)$ accordingly. Interestingly, this expression coincides with the solution given in Eq.~\eqref{easy_sol} for the case $\Omega\leq \frac{\pi}{4}$, meaning in particular that the optimal value of $\Omega$ for $I_\text{anz}$ is given by Eq.~\eqref{eq:bestOmega}. In the Eqs.~\eqref{eq: beta*} and \eqref{eq: I tilde} we can now replace $\tilde I_\text{anz}(\beta;\Omega,q)$ with $I_\text{anz}(\beta;\Omega,q)$, which does not involve any nonlinear optimization.

While we believe this expression to be the true bound, we do not have a formal proof. Anyway, this conjectured expression helps to solve the optimization of interest.

\subsubsection{Certified numerical solution}

As mentioned before, the optimization in Eq.~\eqref{eq: numerical solution} can be easily solved by standard numerical methods. 
However, to provide a strict security guarantee to an actual implementation of DIQKD, such a numerical optimization would need to be done in a certified manner, with a formal proof that the obtained numbers lower bound Eve's conditional entropy on the whole domain. Below we present an algorithm which allows one to do such a certified optimisation based on the Lipshitz continuity of the goal function. The algorithm is rather time-costly, but it only has to be used once the optimal experimental parameters are fixed through an ad hoc maximization of Eq.~\eqref{eq: numerical solution}, cf. below. \\

Concretely, we present in this section an algorithm that approximates the set of possible strategies of Eve, delimited by the bound $\widehat{I}(\beta,\Omega,q)$, from the outside. To avoid the issue of concavity posed by Eq.~\eqref{eq: direct qubit}, we rewrite the problem in the dual form in which we look for the tangent lines
\be\label{eq: dual qubit}\begin{split}
&f(t;\Omega, q) = \max_{{\bf L},\phi} \, H(\rho_E) - H(\rho_E|\hat{\rv{A}_0}) +t\,  \beta_\text{max}({\bf L, } \phi; \Omega)\\
&\beta_\text{max}({\bf L},\phi;\Omega) = \max_{{\bf a}_1, {\bf b}_0,{\bf b}_1} \mathcal{B}_\Omega({\bf L}, \phi, {\bf a}_1, {\bf b}_0, {\bf b}_1),
\end{split}
\ee
to the curve $\widehat{I}(\beta,\Omega,q)$ with different slopes $t$. In Eq.~\eqref{eq: dual qubit} we used the fact that it is only the Bell score that depends on the measurement setting ${\bf a}_1, {\bf b}_0,{\bf b}_1$, so it can be maximized straightforwardly to define $\beta_\text{max}({\bf L},\phi;\Omega)$, see App.~\ref{app: best score} for its closed form expression. \\

Before giving the details on the way we solve this dual form, let us shortly discuss on how it shall be used. We consider an actual implementation of DIQKD with fixed values of $\text{X}_*$,$\text{Y}_*$ and $q_*$ and for which there is an optimal value $\Omega_*$ which saturates the minimum in Eq.~\eqref{eq: best HE}
\be
H(\hat{\rv{A}_0}|E) \geq 1- \widehat{I}(\beta_*;\Omega_*,q_*),
\ee
with $\beta_* = \frac{1}{2}(\C_{\Omega_*}\text{X}_* + \Si_{\Omega_*}\text{Y}_*)$. Therefore, an optimal security guarantee for this particular implementation only requires the knowledge of the function $\widehat{I}(\beta_*,\Omega_*,q_*)$ on a single point. The same lower bond on Eve's conditional entropy can be obtained from the value of the dual bound $f(t;\Omega_*,q_*)$ in Eq.~\eqref{eq: dual qubit} on a single point. Indeed, the concavity of $\widehat{I}(\beta;\Omega, q)$ ensures that there exists a value $t_*$ for which the inequality
\be
f(t_*;\Omega_*,q_*) \geq \widehat{I}(\beta; \Omega_*,q_*) + t_* \beta 
\ee
is saturated at $\beta =\beta_*$ where $f(t_*;\Omega_*,q_*) = \widehat{I}(\beta; \Omega_*,q_*)+t_* \beta_*.$ Using
Eq.~\eqref{eq: dual qubit}, we deduce that
\be\begin{split}
 H(\hat{\rv{A}_0}|E)
&\geq 1- f(t_*;\Omega_*,q_*) + t_*\beta_*\\
& = 1- \widehat{I}(\beta_*;\Omega_*,q_*).
\end{split}
\ee
Hence, for a fixed experimental implementation of the protocol, it is sufficient to certify a single value of the function $f(t_*;\Omega_*,q_*)$ in order to provide a strict and optimal security guarantee. Furthermore, the value $t_*$ is straightforward to find from the knowledge of $\Omega_*,\beta_*, q_*$ and the function $\widehat{I}(\beta; \Omega, q)$. \\

We can now comment on the algorithm to provably upper bound the quantity in Eq.~\eqref{eq: dual qubit}. The basic idea is a branch and bound approach relying on the Lipschitz continuity of the goal function. 
Concretely, we first derive a parametrization of the set $({\bf L},\phi)$ for which the goal function in Eq.~\eqref{eq: dual qubit} is Lipshitz continuous with a constant that we compute. Then, we obtain an upper bound on its value on the whole domain by computing the value of the function on a grid of points. Finally, the algorithm subsequently refines the grid around the points where the value of the function is large in order to approach, step by step, the global maximum. Additional details are given in the next three sections. 

We remark that another algorithm for solving this optimisation was recently developed in a separate work~\cite{Schwonnek20}, which we believe could be adapted to our situation as well.
The basic idea in that work is somewhat different -- for fixed measurement 
angles for Alice and Bob, they find a semidefinite programming (SDP) relaxation for the minimization of Eve's entropy with respect to the state. This SDP is then solved on a grid of angles, and continuity of the goal function with respect to the angles is used to certify that the bound is secure. One advantage of the approach we propose here is that it provably converges to a tight bound, whereas the SDP relaxation in~\cite{Schwonnek20} is not known to be tight\footnote{Note that after the publication of this manuscript, some of the authors have shown how to perform an SDP relaxation that converges to a tight bound as well, see \cite{tan2020improved}.}.
\\

\paragraph{Lipshitz continuity of the entropy with respect to the angle --} A key ingredient in the practical implementation of the desired certified algorithm is that the von Neumann entropy $H(\rho)$ has a bounded rate of change with respect to the angular distance between the two states $\rho$ and $\sigma$ (see App.~\ref{app: Lipsitz entropy}),
\be
A(\rho,\sigma)= \arccos( F(\rho,\sigma)),
\ee
where the fidelity is defined as $F(\rho,\sigma) = \text{tr} |\sqrt{\rho}\, \sqrt{\sigma}|$. This angle is a metric on the set of states~\cite{Nielsen02}. We show that for $n$-dimensional quantum states, the entropy satisfies
\be\label{eq: H slope}
\frac{\left|H(\rho) - H(\sigma)\right|}{A(\rho,\sigma)}<
\begin{cases}
\frac{4\sqrt{r_1(1-r_1)}}{\ln(2)} \sqrt{n-1} & n \leq 4\\
2 \log(n) & n\geq 5.
\end{cases}
\ee
where $r_1 \approx 0.203$ (see Eq.~\eqref{eq:r1value} below for details).
This contrasts with the trace distance, another metric on the set of quantum states, for which the entropy has infinite rate of change around non-full-rank states.\\

\paragraph{Bounding the gradient of the goal function--} To apply the continuity bound above to the goal function in Eq.~\eqref{eq: dual qubit}, we use the following parametrization of the state
\be\label{eq: L with angles}
\sqrt{{\bf L}} = 
\left(\begin{array}{c}
\cos(\alpha) \cos(\mu)\\
\cos(\alpha) \sin(\mu)\\
\sin(\alpha) \cos(\xi)
\\
\sin(\alpha) \sin(\xi)
\end{array}
\right),
\ee
so that the quantum models are described by four angles
\be
{\bf x}= (\alpha, \mu, \xi, \phi),
\ee
with ${\bf x} \in [0,\frac{\pi}{4}]^{\times 3}\times [0, \frac{\pi}{2}]$. Here, the condition $\mu,\xi \leq \frac{\pi}{4}$ follow from $L_1\geq L_2$, $L_3\geq L_4$. The bound $\alpha\leq\frac{\pi}{4}$ is a consequence of $L_1+L_2\geq L_3+L_4$, which can be imposed on the states as an alternative to $L_1-L_2\geq L_3-L_4$, see App.~\ref{app: parametrization}.\\

To obtain a bound on the gradient of the goal function 
\be
G({\bf x}) = H({\bf L}) - H(\rho_E|\hat{a}=+1) +t\,  \beta_\text{max}({\bf L, } \phi; \Omega)
\ee
we bound the gradient of each term independently, as described in App.~\ref{app: continuity of goal} and \ref{app: gradient of goal}. For the entropic terms, we use Eq.~\eqref{eq: H slope}, while the computation of the maximal gradient is quite straightforward for the CHSH score. Combining the three terms, we obtain a general bound
\be
|\nabla_{\bf x} G| \leq  12.7 + 7\,  t
\ee
on the whole domain of $\bf x$. \\

With this global bound on the gradient, a certified maximization of the function $G({\bf x})$ can be obtained with a branch and bound approach. In order to do so, we extended the code developed in Ref.~\cite{Valcarce2020} to include global optimization. A detailed description of this code can be found in the App.~\ref{app: algo}. A python implementation as well as an example for our case of interest, can be found on Gitlab\footnote{\url{https://gitlab.com/plut0n/bcert}}.\\

\section{Results}

\begin{figure}[h]
    \centering
    \includegraphics[width=0.5\textwidth]{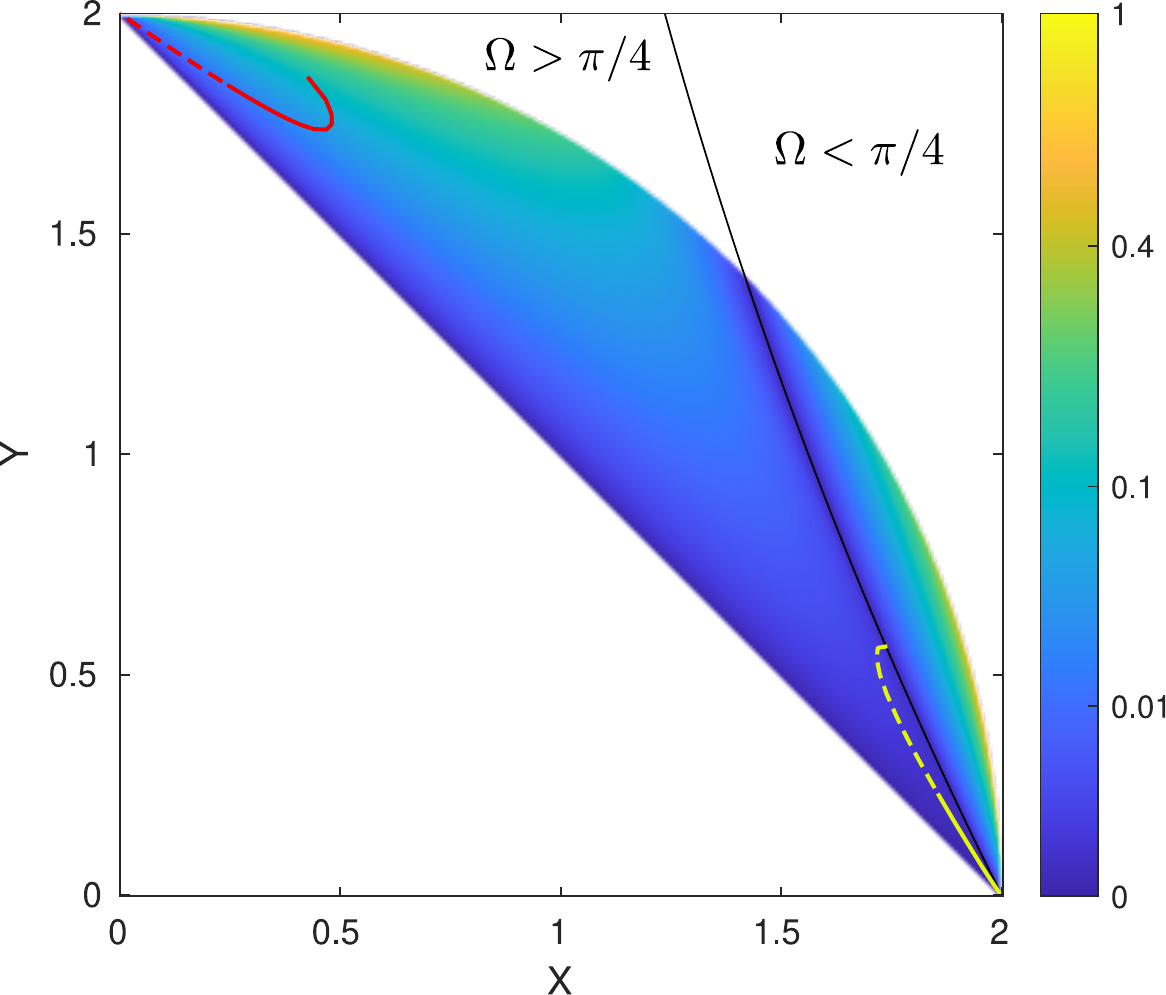}
    \caption{Difference between the bound on Eve's conditional entropy $H(\widehat{\rv{A}}_0|E)$ computed as a function of X and Y, and as a function of the CHSH quantity $S=\text{X}+\text{Y}$, for $p=0$. In the presence of noisy preprocessing (i.e. $p>0$), the advantage follows a similar distribution, but is smaller in magnitude. The CHSH bound is only optimal along the curve $\text{X}(\text{X}+\text{Y})=4$. The advantage on the right-hand side of this curve is obtained with $\Omega<\pi/4$, and on the left-hand side with $\Omega>\pi/4$. The yellow (red) curve shows the trajectory in the X-Y plane which optimizes the key rate in an optical implementation of DIQKD for low (high) detection efficiencies. At the efficiency $\eta=0.923$, it is better to switch from one curve to the other one (at the transition between full and dashed curves). The points with $\Omega>\pi/4$ were computed both with the heuristic method and with the ansatz described in Sec.~\ref{sec: complicated}.}
    \label{fig:advantageHAE}
\end{figure}

\paragraph{Improved bound on Eve's conditional entropy--}
In order to demonstrate the advantage of considering the pair of variables $(\text{X},\text{Y})$ when bounding Eve's conditional entropy, we compute the bound on $H(\widehat{\rv{A}}_0|E)$ for different values of X, Y and $p$ and compare it to the bound obtained from the CHSH score~\cite{Ho20}. Because of the two regimes identified earlier, we compute both the optimal bound assuming $\Omega\leq \pi/4$ and the one assuming $\Omega > \pi/4$ for each value of X and Y, and keep the best one. In the case $\Omega\leq \pi/4$, the optimal choice of $\Omega$ is readily given by Eq.~\eqref{eq:bestOmega}. In the other case, we optimize the bound over $\Omega\in(\pi/4,\pi/2]$. The difference between this optimal bound on $H(\widehat{\rv{A}}_0|E)$ given $X$ and $Y$ and Eq.~\eqref{formula_pironio} is shown in Fig.~\ref{fig:advantageHAE}. It shows that our bound on $H(\widehat{\rv{A}}_0|E)$ is better than the one derived from the CHSH score, except along the line satisfying $\text{X}(\text{X}+\text{Y})=4$. We emphasise that our derivation of the bound is constructive, that is we find the optimal attack that gives $H(\widehat{\rv{A}}_0|E)$ to Eve. The final bound is thus tight, up to the precision of the numerical algorithms for $\Omega > \frac{\pi}{4}$.\\ 

\paragraph{Implication for a practical realization of DIQKD--}
We now study the potential impact of our bound on practical realizations of DIQKD. In the limit of asymptotically many repetitions, an implementation is uniquely characterized by its key rate $r,$ which is given in Eq.~\eqref{keyrate}. In our case, this key rate is determined from three quantities: $\text{X}$, $\text{Y}$ and Bob's uncertainty about Alice's key generating bit as a function of the noisy preprocessing parameter, given by $H(\widehat{\rv{A}}_0|\rv{B}_2)$. Hence, in order to find the optimal design for an experimental implementation of DIQKD, we express the quantities
\be
\text{setup} \simeq \big(\text{X}, \text{Y}, H(\hat{\rv{A}_0}|\rv{B}_2) \big)
\ee
as a function of the model's parameters. Then, we maximize the key rate over these parameters:
\be\small\begin{split}
\check{r}&= \max_{\text{setup},\Omega, q} H(\widehat{\rv{A}}_0|E)-H(\widehat{\rv{A}}_0|\rv{B}_2) \\
&=\max_{\text{setup},\Omega, q} 1 - \widehat{I}\left(\frac{\C_\Omega \text{X} + \Si_\Omega \text{Y}}{2};\Omega,q \right)-H(\hat{\rv{A}_0}|\rv{B}_2)
\end{split}
\ee
Solving this maximization gives a bound on the key rate $\check{r}$, the values $\big(\text{X}_*, \text{Y}_*, H(\widehat{\rv{A}}_0|\rv{B}_2) \big)$ expected for a given implementation, as well as the optimal values of the parameters $\Omega_*$ and  $q_*$ for this implementation.\\

\paragraph{SPDC-based implementation of DIQKD--}
Photonic experiments using a source based on spontaneous parametric down conversion (SPDC) are one of the most promising setups for implementing DIQKD, as shown by recent experiments reporting on the violation of a Bell inequality without the fair sampling assumption~\cite{Giustina13, Christensen13, Shalm15, Giustina15, Shen18, Liu18}. We consider such a setup in which an SPDC source is used to create and distribute polarization entanglement between distant parties who perform measurements as requested here in the proposed protocol. The main limitation in this setup is the overall detection efficiency, i.e. the possibility of losing a photon at any point between its creation at the source and its final detection. To reflect photon losses and non-unit detection efficiency, the transmission channel between the source and the parties is modeled as a lossy channel with an overall transmission $\eta$. It is also important to include the statistics of an SPDC source which does not produce a two-qubit state, but a state that contains vacuum and multiple photon components. We invite the reader to look at Ref.~\cite{Ho20} to get explicit expressions of the exact statistics created by this source as well as a description of tunable parameters. \\

When computing the key rate for an SPDC source with a security determined by the CHSH score, any values of X and Y with the same sum impose the same bound on Eve's conditional entropy $H(\widehat{\rv{A}}_0|E)$. In a Eberhard-like scenario where a significant fraction of the entangled particles can be lost before yielding their measurement result, it is advantageous for Alice to use two measurement settings with different overlap with her Schmidt basis in order to maximize the CHSH quantity~\cite{Eberhard93}. It is then easier for Bob to guess the outcome of one of the two measurements (the one best aligned with the Schmidt basis). When the key rate is extracted from this measurement to minimize the cost of error correction (see Ref.~\cite{Ho20} for more details), the value of the X quantity is then larger than Y, as shown in Fig.~\ref{fig:advantageHAE}. But in this case, the (X,Y) values are very close to the line $\text{X}(\text{X}+\text{Y})=4$, for which there is no advantage. Therefore, we only expect a small improvement in the key rate here.\\

Given the values shown in Fig.~\ref{fig:advantageHAE}, a better bound on Eve's information would be obtained if the same CHSH value was obtained with the contributions from X and Y being inverted, i.e. with $\text{Y}>\text{X}$. However, this requires Alice to define her key-generating measurement as being the one less aligned with the Schmidt basis, hence leading to an increase of the error correction cost. In this case, both conditional entropies $H(\widehat{\rv{A}}_0|E)$ and $H(\widehat{\rv{A}}_0|{\rv B}_2)$ are larger, and we need to check which one increases the most in order to infer a possible gain on the key rate. As it turns out, the tradeoff between these two entropies increase depends on the detection efficiency.\\

\begin{figure}
    \centering
    \includegraphics[width=0.5\textwidth]{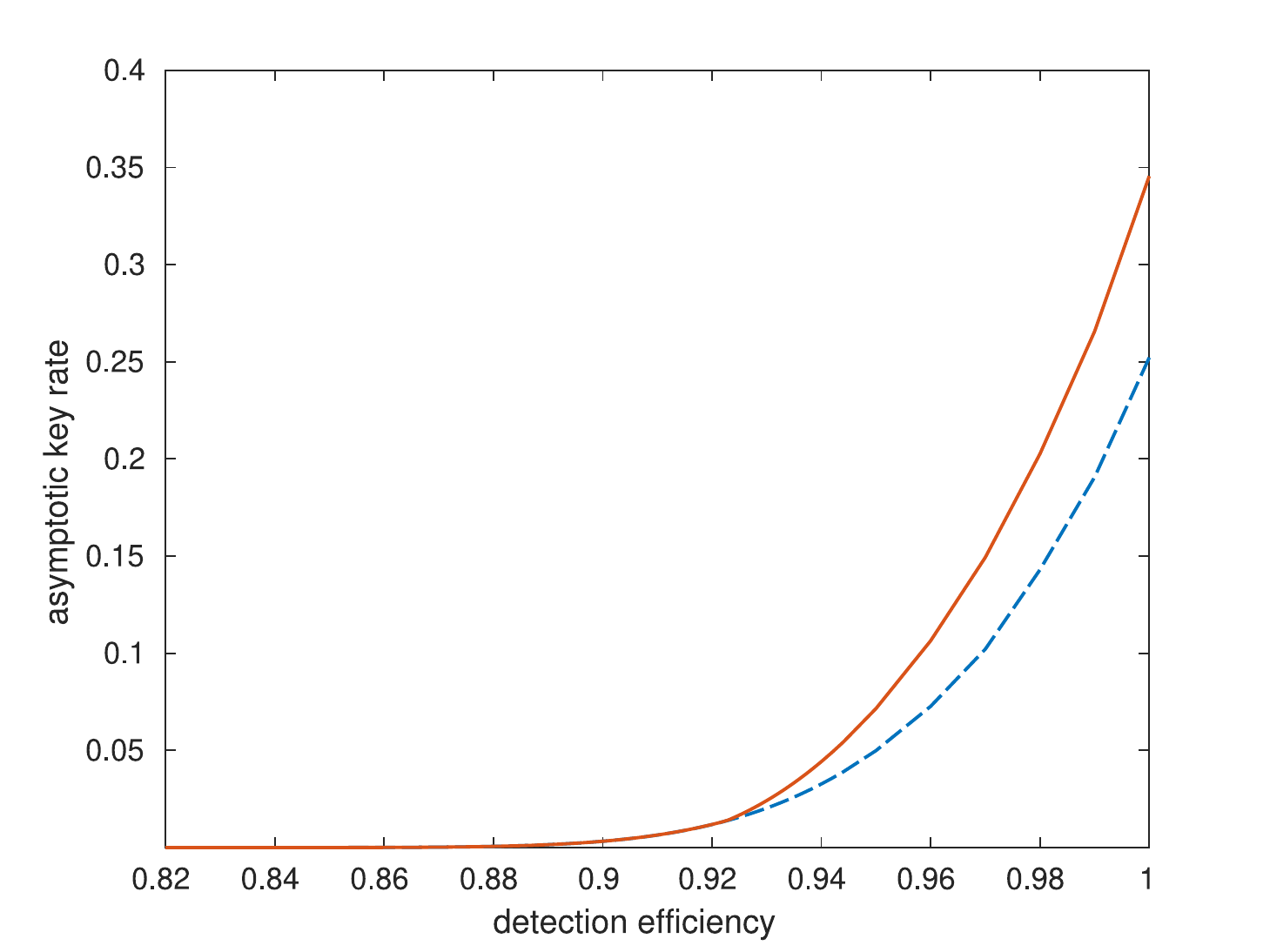}
    \caption{Key rate achievable with a photonic setup as a function of the symmetric detection efficiency $\eta$. The dashed blue curve corresponds to the key rate achieved with noisy preprocessing and a security based on the CHSH quantity alone; it is nonzero for efficiencies above $\sim$0.828~\cite{Ho20}. The continuous red curve bases its security on the values of both X and Y (and also includes noisy preprocessing). A significant increase of the key rate is possible for efficiencies above $\eta > 0.923$. At $\eta=1$, the obtained key rate is 0.346 instead of 0.252.}
    \label{fig:keyrate}
\end{figure}

Namely, there are two regimes, as shown in Fig.~\ref{fig:keyrate}. When the detection efficiency is larger than $\sim 0.923$, relabelling the measurements in order to enter the region with $\text{Y}>\text{X}$ (where inequalities with $\Omega>\pi/4$ significantly improve the bound on $H(\widehat{\rv{A}}_0|E)$) is advantageous, because the increase in Bob's uncertainty $H(\widehat{\rv{A}}_0|\rv{B}_2)$ is smaller. The red curve in Fig.~\ref{fig:advantageHAE} shows the corresponding trajectory in the X-Y plane. When $\eta < 0.923$, the cost of error correction become prohibitive compared to the potential increase in Eve's uncertainty, and it is better to stay in the region $\text{X}>\text{Y}$, as represented by the yellow curve in Fig.~\ref{fig:advantageHAE}. There, a small increase of the key rate is still found because the correlations do not satisfy $\text{X}(\text{X}+\text{Y})=4$ exactly. However, this condition is only slightly violated, resulting in an increase in key rate smaller than $\sim 10^{-4}$, which is practically negligible. The critical detection efficiency then also remains at $\sim 83\%$, unchanged compared to a bound based on CHSH alone.\\

\paragraph{Comparison of qubit vs SPDC bounds--}
To illustrate the impact of the photon statistics of SPDC sources on DIQKD, we now consider a simpler model in which the state shared between Alice and Bob is a two-qubit state:
\begin{equation}
    \ket{\psi}=\cos(\theta)\ket{00}+\sin(\theta)\ket{11}.
\end{equation}
We are not aware of physical setups allowing one to produce a state with $\theta=\pi/4$ but not allowing for a different value of $\theta$. Still, for the sake of the discussion, we distinguish between the cases where the state can be either constrained to be maximally entangled, i.e. $\theta=\pi/4$, or can have an arbitrary parameter $\theta\in[0,\pi/4]$.\\

\begin{table}[]
    \centering
{\small    \begin{tabular}{c|ccc}
        Keyrate formula & Singlet & Qubit & SPDC \\
        \hline
        $1-I(\beta;\frac{\pi}4,0) - h(Q)$ & 0.923 & 0.893 & 0.927 \\ 
        $1-I(\beta;\frac{\pi}4,0) - H(\rv{A}_0|\rv{B}_2)$ & 0.908 & 0.865 & 0.909 \\ 
        $1-I(\beta;\frac{\pi}4,q)-H(\rv{A}_0|\rv{B}_2)$ & 0.903 & 0.826 & 0.826 \\ 
        $1-I(\beta;\Omega,q) - H(\rv{A}_0|\rv{B}_2)$ & 0.900 & 0.826 & 0.826\\ 
    \end{tabular}\\
}
    \caption{Critical detection efficiencies for various states and protocols. Here, $Q$ is the quantum bit error rate (QBER). These thresholds are compared in Fig.~\ref{fig:SPDCvsQubit}.}
    \label{tab:thresholds}
\end{table}

\begin{figure}
    \centering
    \includegraphics[width=0.5\textwidth]{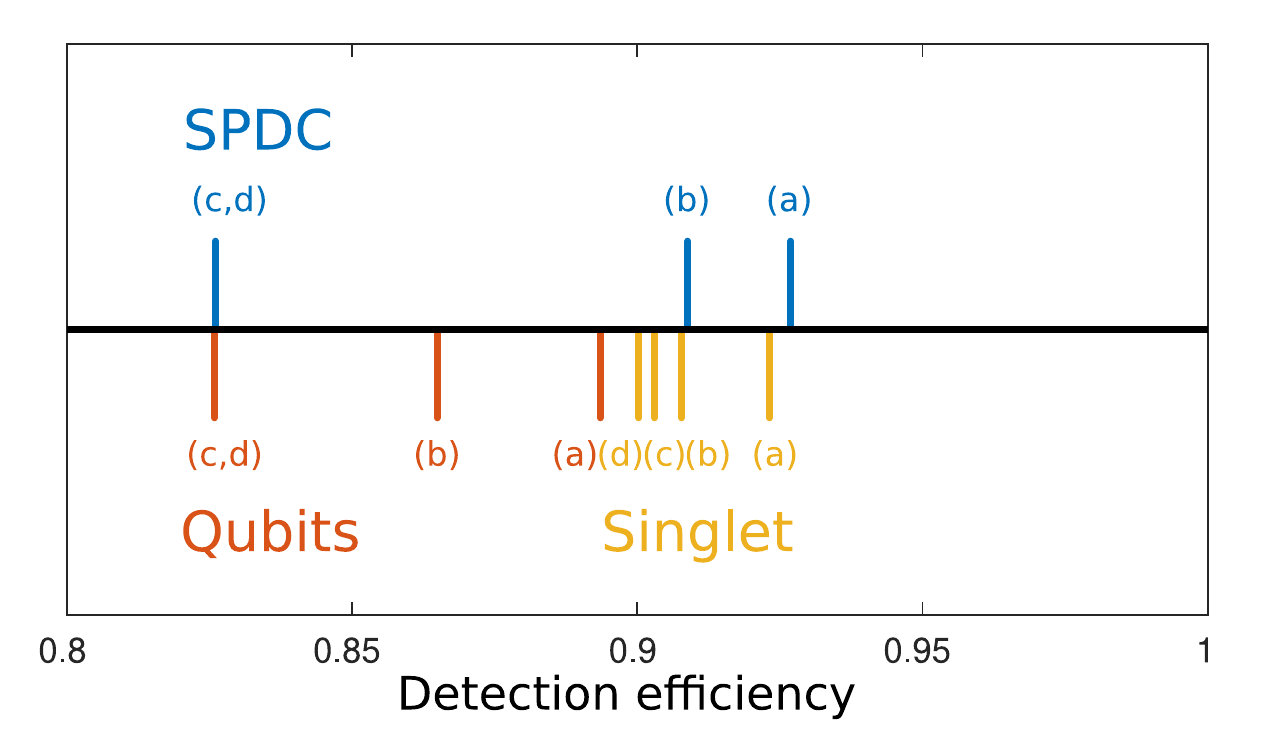}
    \caption{Comparison of the critical detection efficiencies for several setup models. Blue markers are for SPDC statistics, yellow ones for measurement on a maximally entangled two-qubit state, and red ones for measurement on an arbitrary two-qubit state. The four protocols of Tab.~\ref{tab:thresholds} are considered here: the security of (a) is based on the CHSH value following Eq.~\eqref{formula_pironio} and error correction based on the QBER~\cite{Pironio09}, (b) Eq.~\eqref{formula_pironio} with error correction based on conditional entropy $H(A|B)$~\cite{Ma12}, (c) security from CHSH with noisy preprocessing and error correction based on $H(A|B)$ following~\cite{Ho20}, (d) security from X and Y with noisy preprocessing and error correction based on $H(A|B)$.}
    \label{fig:SPDCvsQubit}
\end{figure}

In Tab.~\ref{tab:thresholds}, see also Fig.~\ref{fig:SPDCvsQubit}, we report the critical detection efficiencies corresponding to various security analyses applied on these implementations. Like for the SPDC model, no advantage on the critical detection efficiency is found when using arbitrary two-qubit systems. A small advantage is however present when restricting to measurements on the singlet state. Still, this model is not optimal and it remains of course better to use partially entangled states. In fact, even partially entangled states produced by an SPDC source perform better.\\

In this respect, it is worth noticing here that the performance of an SPDC source is essentially comparable to that of an arbitrary two-qubit state once noisy preprocessing is taken into consideration, i.e. the requirement on the detection efficiency is very similar. In the case without noisy preprocessing, the state produced by these physical sources do not have a better tolerance to losses than measurements on a maximally entangled state. This suggests that noisy preprocessing is a key ingredient for a first proof of principle implementation with an SPDC source~\cite{Ho20}.

\section{Discussion}
In this paper, we introduced a refinement of the usual CHSH-based analysis of DIQKD experiments: Instead of projecting the measurement statistics onto a single line giving the CHSH score $\text{X} +\text{Y}$, we kept the information about the individual values of X and Y throughout the whole security analysis. We found that this refined analysis gives a more restrictive bound on the information available to the eavesdropper for almost all values X and Y in the quantum set.\\
 
When applying our results to photonic implementations of DIQKD with a SPDC source, we found that the key rate is improved by a factor going up to 37\% for unit detection efficiency. On the other hand, we could not find any improvement for the critical detection efficiency as compared to the CHSH protocol with noisy processing presented in Ref.~\cite{Ho20}. However, we focused on a given photonic implementation, and the question of the most favorable optical setup combining squeezing operations, displacement operations, linear optical elements and photon counting techniques is still open. Advanced techniques using automated design of quantum experiments based on reinforcement learning which already proved to be useful to optimize the CHSH score~\cite{melnikov2020setting} are inspiring. Applying them to the proposed protocol in order to reduce the required detection efficiency for implementing DIQKD appears to be promising for future work.\\

Finally, we would like to remark that the certified numerical techniques we proposed also open up the possibility of bounding Eve's information reliably when more correlators, or even the full measurement statistics, are taken into account.

\section{Note added}
While writing this manuscript, we became aware of another manuscript~\cite{Woodhead20} reporting on similar results.

\section{Acknowledgments}
 We thank Stefano Pironio for pointing out that the convexity argument provided in~\cite{Ho20} was not complete, see the discussion in Sec.~\ref{sec: simple case} and App.~\ref{app: convex}. We acknowledge funding by the Swiss National Science Foundation (SNSF), through the Grants PP00P2-179109 as well as via the National Center for Competence in Research for Quantum Science and Technology (QSIT). 

\printbibliography

\appendix

\begin{widetext}

\section{Analytical results}

\subsection{Parametrization of two-qubit models}
\label{app: parametrization}

Following the logic of \cite{Pironio09}, we assume, without loss of generality that after the qubit reduction, the state shared by Alice, Bob and Eve is of the form
\be
\ket{\Psi}_{ABE} = \sum_{i=1}^4 \sqrt{L_i} \ket{\Phi^i}_{AB}\ket{i}_E
\ee
where we $\ket{\Phi_i}=\{ \Phi^+, \Psi^-,\Phi^-,\Psi^+\}_{i=1}^4$, with nonnegative $L_1\geq L_2$ and $L_3\geq L_4$, and the measurements $A_x$, $B_y$ appearing in the constraint are in the x-z plane
\be
A_x = {\bf a}_x^T \binom{\sigma_z}{\sigma_x} \qquad 
B_y = {\bf b}_y^T \binom{\sigma_z}{\sigma_x}.
\ee
The key generating setting is explicitly parametrized by an angle $\phi$
\be
{\bf a}_0=\binom{\C_\phi}{\Si_\phi}.
\ee
(As mentioned in the main text we use a compact notation $\C_\phi = \cos(\phi)$ and $\Si_\phi = \sin(\phi)$ throughout the paper.) One notes that the application of a unitary transformation $\sigma_Z$ on Alice's system is equivalent to changing the state and the measurements as
\be\begin{split}
    (L_1,L_2,L_3,L_4) &\to (L_3,L_4,L_1,L_2)\\
    {\bf a}_x &\to \sigma_z {\bf a}_x\\
    {\bf b}_y &\to {\bf b}_y.
\end{split}
\ee
The two situations are completely equivalent for our purpose, so our parametrization of quantum models is actually redundant. To avoid this, we can introduce an order relation between the pairs of coefficients $(L_1,L_2)$ and $(L_3,L_4)$ (permuted by a basis transformation). In particular, we will impose
\be
L_1-L_2 \geq L_3-L_4
\ee
below, but will also use $L_1+L_2\geq L_3+L_4$ for the certified numerical algorithm.

We also introduce a different parametrization of the probability simplex $\bf L$ with a vector ${\bf T} = (T_z, T_x, T_p)$ given by
\be
\begin{cases}
 T_z = (L_1-L_2)+(L_3-L_4)\\
 T_x = (L_1 -L_2) -(L_3-L_4)  \\
 T_p = L_1 +L_2-L_3-L_4.
\end{cases}
\ee
The conditions $L_1\geq L_2, L_3\geq L_4$ and $L_1-L_2\geq L_3-L_4$ enforce
\be\begin{split}
    0\leq T_x &\leq T_z\leq 1\\
T_z+T_x-1\leq &\ T_p \leq 1-(T_z-T_x).
\end{split}
\ee

\subsection{Entropies of $\rho_{E|\hat{a}=\pm 1}$}
\label{app: entropy formulas}

Having introduced a parametrization of quantum models, we now express the quantities of interest $H(\rho_E), H(\rho_E|\hat{\rv{A}_0})$ and $B_\Omega$ as functions of the distribution $\bf L$ describing the state $\ket{\Psi}_{ABE}$ and the measurement settings ${\bf a}_x, {\bf b}_y$. The marginal state of Eve is straightforward to write down:
\be
\rho_E =\left(\begin{array}{cccc}
L_1 &&&\\
& L_2 && \\
&& L_3 & \\
&&& L_4
\end{array}\right).
\ee
For the conditional states, we have
\be\rho_{E|\hat{a}=+1}= 2\, \text{tr}_{AB} \frac{1+\hat{A}_0}{2}\otimes \Id_{BE}\prjct{\Psi}_{ABE}=
\left(
\begin{array}{cccc}
 L_1 & 0 & \C_\phi \sqrt{L_1 L_3 q} & \Si_\phi \sqrt{L_1 L_4 q} \\
 0 & L_2 & \Si_\phi \sqrt{L_2 L_3 q} & -\C_\phi \sqrt{L_2 L_4 q} \\
 \C_\phi \sqrt{L_1 L_3 q} & \Si_\phi \sqrt{L_2 L_3 q} & L_3 & 0 \\
 \Si_\phi \sqrt{L_1 L_4 q} & -\C_\phi \sqrt{L_2 L_4 q} & 0 & L_4 \\
\end{array}
\right),
\ee
where the factor of two arises because the probability to observe the outcome $\hat{\rv{A}_0}=1$ is simply $\frac{1}{2}$. The conditional state for the other outcome $\hat{\rv{A}_0} =-1$ can be easily obtained by noticing that the two outcomes are interchanged by a mapping $\hat{A}_0 \to - \hat{A}_0$, i.e. with the inversion of the vector $\binom{\C_\phi}{\Si_\phi}\to -\binom{\C_\phi}{\Si_\phi}.$ Consequently, we have
\be
\rho_{E|\hat{a}=-1}=
\left(\begin{array}{cccc}
     1&&&  \\
     & 1 &&\\
     &&-1&\\
     &&&-1
\end{array}\right)
\rho_{E|\hat{a}=1}
\left(\begin{array}{cccc}
     1&&&  \\
     & 1 &&\\
     &&-1&\\
     &&&-1
\end{array}\right).
\label{eq: change outcomes}
\ee
It follows that the entropies of both conditional states are equal and
\be
H(\rho_E|\hat{\rv{A}_0}) = \frac{1}{2} H(\rho_{E|\hat{a}=+1})+ \frac{1}{2} H(\rho_{E|\hat{a}=-1}) = H(\rho_{E|\hat{a}=+1}).
\ee
Analogously to Eq.~\eqref{eq: change outcomes}, there are unitary transformations (with different positions of $1$ and $-1$ on the diagonal) that correspond to the transformations $\binom{\C_\phi}{\Si_\phi}\to \binom{-\C_\phi}{\Si_\phi}$ and $\binom{\C_\phi}{\Si_\phi}\to \binom{\C_\phi}{-\Si_\phi}$. This implies that $H(\rho_E|\hat{\rv{A}_0})$ and $H(\hat{\rv{A}_0}|E)$ do not depend on these sign changes, i.e. they are only functions of $\C_\phi^2$ and $\Si_\phi^2$.\\

\subsection{Optimal key generating setting ${\bf a}_0$}
\label{app: best phi}

In this section, we give the minimal angle $\phi$ for which constraint $\mathcal{B}_\Omega \geq \beta$ can be fulfilled for a given $\bf L$. We first focus on the possible values of $\phi.$ The expected Bell score is straightforward to compute:
\be\begin{split}
    \mathcal{B}_\Omega &= \frac{1}{2}\text{tr} \left( \left(\left(\C_\Omega A_0\otimes(B_0+B_1) +\Si_\Omega A_1\otimes(B_0-B_1)\right)\otimes \Id_E \right) \prjct{\Psi}_{ABE} \right)\\
    &=\frac{1}{2} \C_\Omega \, {\bf a}_0^T 
    \left(\begin{array}{cc}
         T_z&  \\
         & T_x
    \end{array}\right) ({\bf b}_0 +{\bf b}_1)
    +\frac{1}{2} \Si_\Omega \, {\bf a}_1^T 
    \left(\begin{array}{cc}
         T_z&  \\
         & T_x
    \end{array}\right) ({\bf b}_0 -{\bf b}_1)
\end{split}
\ee
From this expression we notice that any of the following transformations of the key generating setting: 
\be
{\bf a}_0 \to \left(\begin{array}{cc}
(-1)^{s_1} & \\
& (-1)^{s_2}
\end{array}\right) {\bf a}_0,
\ee
with $s_1, s_2 \in \{0,1\}$ can be compensated by applying the same transformation to the remaining settings ${\bf a}_1$, ${\bf b}_y$ to give the same Bell score $\mathcal{B}_\Omega$. Furthermore, we have seen that this transformation does not change Eve's conditional entropy $H(\hat{\rv{A}_0}|E)$. Therefore we can always restrict ${\bf a}_0$ to the positive quadrant of the circle
\be
\phi \in [0,\frac{\pi}{2}]
\ee
without loss of generality.
We now express the Bell score with Bob's settings parametrized as in Eq.~\eqref{eq: B0B1 param}:
\be
\mathcal{B}_\Omega = \left\langle \C_\Omega \C_\theta\,  {\bf a}_0 \cdot \binom{Z}{X}  \otimes {\bf c}\cdot \binom{Z}{X} +\Si_\Omega \Si_\theta\, {\bf a}_1 \cdot \binom{Z}{X} \otimes {\bf c}_\perp\cdot \binom{Z}{X} \right\rangle.
\ee
Computing the expected value of the operators on our Bell diagonal state gives
\be \label{eq: Bell score full}
\mathcal{B}_\Omega  =  \C_\Omega \C_\theta\,  {\bf a}_0^T  
\left(\begin{array}{cc}
T_z & \\
& T_x \\
\end{array}\right)
{\bf c} +
 \Si_\Omega \Si_\theta \, {\bf a}_1^T
\left(\begin{array}{cc}
T_z & \\
& T_x \\
\end{array}\right)
{\bf c}_\perp.
\ee
We introduce an angle $\gamma$ to parametrize the vectors ${\bf c}$ and ${\bf c}_\perp$:
\be
{\bf c} = \binom{\C_\gamma}{\Si_\gamma}, \quad {\bf c}_\perp = \binom{-\Si_\gamma}{\C_\gamma}.
\ee
The maximization with the second setting of Alice is straightforward, that is
\be
\max_{{\bf a}_1}\,   \Si_\theta {\bf a}_1^T
\left(\begin{array}{cc}
T_z & \\
& T_x \\
\end{array}\right)
{\bf c}_\perp = \|   \Si_\theta
\left(\begin{array}{cc}
T_z & \\
& T_x \\
\end{array}\right)
{\bf c}_\perp\| =|\Si_\theta|\sqrt{T_z^2 \Si_\gamma^2 + T_x^2 \C_\gamma^2}.
\ee
The Bell score (optimized with respect to ${\bf a}_1$) becomes
\be
\mathcal{B}_\Omega = \C_\Omega \C_\theta\, \binom{\C_\phi}{\Si_\phi}^T\!\!  \left(\begin{array}{cc}
T_z & \\
& T_x \\
\end{array}\right) 
\binom{\C_\gamma}{\Si_\gamma} +  \Si_\Omega |\Si_\theta|\sqrt{T_z^2 \Si_\gamma^2 + T_x^2 \C_\gamma^2}.
\ee
Since $T_z,T_x\geq0$ and $\phi,\Omega\in[0,\frac{\pi}{2}]$, we can also restrict the angle $\theta$ and $\gamma$ to the interval $[0,\frac{\pi}{2}]$ without loss of generality, and drop the absolute value: $|\Si_\theta|=\Si_\theta$. The constraint $\mathcal{B}_\Omega\geq \beta$ takes the form
\be
\C_\Omega \, \binom{\C_\phi}{\Si_\phi}^T\!\!  \left(\begin{array}{cc}
T_z & \\
& T_x \\
\end{array}\right) 
\binom{\C_\gamma}{\Si_\gamma} \geq \frac{\beta - \Si_\Omega \Si_\theta\sqrt{T_z^2 \Si_\gamma^2 + T_x^2 \C_\gamma^2}}{\C_\theta}.
\label{eq: constraint frac}
\ee 
Recall that we wish to find the minimal $\phi$ for which this inequality can be fulfilled for at least one value of the free parameters $\theta$ and $\gamma$.
We observe that if the right hand side (RHS) can become zero or negative by some choice of $\theta$ and $\gamma$, the constraint becomes trivial. Since $T_z\geq T_x$, this is possible for
\be
\beta^2 \leq \Si_\Omega^2 T_z^2  \implies \phi_*=0.
\ee
\\

In the following we assume that this is not the case, i.e.~$\beta^2 > \Si_\Omega^2 T_z^2$. The angle $\theta$ only appears on the right of the inequality, so to satisfy the inequality our best choice is to minimize the RHS with respect to $\theta$. The expression
\be
 \frac{\beta - \Si_\Omega \Si_\theta\sqrt{T_z^2 \Si_\gamma^2 + T_x^2 \C_\gamma^2}}{\C_\theta}
\ee
either has a local minimum that can be found by setting its derivative to zero, or there is no local minimum and the expression is minimal at the boundary $\theta=0$ since it diverges for $\theta \to \frac{\pi}{2}.$ Differentiating the expression with respect to theta, we find that a local minumum does exist at
\be
\Si_\theta =\frac{ \Si_\Omega\sqrt{T_z^2 \Si_\gamma^2 + T_x^2 \C_\gamma^2}}{\beta},
\ee
(recall the assumption above). Plugging this value into Eq.~\eqref{eq: constraint frac} allows us to rewrite the constraint as
\be
 \C_\Omega \, \binom{\C_\phi}{\Si_\phi}^T\!\! \left(\begin{array}{cc}
T_z & \\
& T_x \\
\end{array}\right)
\binom{\C_\gamma}{\Si_\gamma}\geq \sqrt{\beta^2 - \Si_\Omega^2(T_z^2 \Si_\gamma^2 + T_x^2 \C_\gamma^2)},
\ee
which we rewrite as
\be
\binom{\C_\phi}{\Si_\phi}^T {\bf v}_\gamma \geq 1, \qquad \text{with} \qquad {\bf v}_\gamma = \frac{\C_\Omega}{\sqrt{\beta^2 - \Si_\Omega^2(T_z^2 \Si_\gamma^2 + T_x^2 \C_\gamma^2)}} \binom{T_z \C_\gamma}{T_x \Si_\gamma}.
\ee

Now it becomes simple to check if the constraint can be satisfied at all. The vector ${\bf v}_\gamma$ belongs to the positive quadrant of the plane with ${\bf v}_0 \parallel \binom{1}{0}$ and ${\bf v}_{\pi/2} \parallel \binom{0}{1}$, hence the inequality ${\bf a}_0 \cdot {\bf v}_\gamma \geq 1$ can be satisfied if and only if the length of the vector ${\bf v}_\gamma$ reaches 1, i.e. 
\be
|{\bf v}_\gamma|^2= \frac{\C_\Omega^2(T_z^2\C_\gamma^2 + T_x^2 \Si_\gamma^2)}{\beta^2 - \Si_\Omega^2(T_z^2 \Si_\gamma^2 + T_x^2 \C_\gamma^2)}\geq 1
\ee
for some $\gamma$.
We rewrite this inequality as
\be
\frac{1}{2} \left( T_z^2 + T_x^2 + (T_z^2-T_x^2)\C_{2\gamma}\C_{2\Omega} - 2\beta^2\right) \geq 0.
\ee
Given that $T_z^2\geq T_x^2$, the left hand side (LHS) is maximal for $\gamma=0$ if $\Omega\leq \frac{\pi}{4}$ and for $\gamma=\frac{\pi}{2}$ if $\Omega > \frac{\pi}{4}$. Hence, $\mathcal{B}_\Omega \geq \beta$ can be fulfilled if and only if
\be\label{eq: constraint final}
\boxed{
\begin{split}
&\text{for} \, \Omega\leq \frac{\pi}{4}, \qquad \C_\Omega^2 T_z^2 + \Si_\Omega^2 T_x^2 \geq \beta^2\\
&\text{for} \, \Omega > \frac{\pi}{4}, \qquad \Si_\Omega^2 T_z^2 + \C_\Omega^2 T_x^2 \geq \beta^2.
\end{split}}
\ee\\

Consider the first case $\boxed{\Omega\leq \frac{\pi}{4}}$. Setting $\gamma =0$ one verifies  that if the constraint can be fulfilled, then it can be fulfilled with $\phi =0$:
\be
\binom{1}{0}^T {\bf v}_0 = \frac{\C_\Omega T_z}{\sqrt{\beta^2 - \Si_\Omega^2 T_x^2}} \geq 1 \ \ \iff\ \  \C_\Omega^2 T_z^2 + \Si_\Omega^2 T_x^2 \geq \beta^2.
\ee
Hence, in this case the minimal possible angle is $\phi_*=0$ as long as the Bell score can be attained as formalized by Eq.~\eqref{eq: constraint final}.\\

The other case $\boxed{\Omega> \frac{\pi}{4}}$ is less trivial. Assume that the constraint can be satisfied,  $\Si_\Omega^2 T_z^2 + \C_\Omega^2 T_x^2 \geq \beta^2$. We first check if $\phi=0$ is a solution
\be
\left(\binom{1}{0}^T {\bf v}_\gamma \right)^2= \frac{\C_\Omega^2 \C_\gamma^2 T_z^2}{\beta^2 - \Si_\Omega^2(T_z^2 \Si_\gamma^2 + T_x^2 \C_\gamma^2)} \geq 1 \ \ \iff\ \  \C_\gamma^2 (\C_\Omega^2 T_z^2 + \Si_\Omega^2 T_x^2 -\Si_\Omega^2 T_z^2) +  \Si_\Omega^2 T_z^2  \geq \beta^2.
\ee
As $\Si_\Omega^2 T_z^2 + \C_\Omega^2 T_x^2 \geq \beta^2\geq \Si_\Omega^2 T_z^2$, the LHS is maximal for $\gamma=0$. Therefore $ \phi_*=0$ iff
\be
\C_\Omega^2 T_z^2 + \Si_\Omega^2 T_x^2 \geq \beta^2.
\ee
We now assume 
\be
\Si_\Omega^2 T_z^2 + \C_\Omega^2 T_x^2 \geq \beta^2 > \C_\Omega^2 T_z^2 + \Si_\Omega^2 T_x^2,
\ee
such that the constraint can be fulfilled but not with $\phi=0$. To find the minimal $\phi$ that allows to do so we look on the dependence of the length of the vector ${\bf v}_\gamma$  on $\gamma$. We compute 
\be
\frac{d}{d\gamma}|{\bf v}_\gamma|^2 = \frac{(T_z^2-T_x^2)\C_\Omega^2\Si_{2\gamma}}{(\beta^2- (T_z^2 \Si_\gamma^2+T_x^2 \C_\gamma^2)\Si_\Omega^2)^2}((T_z^2+T_x^2)\Si_\Omega^2 - \beta^2),
\ee
here $(T_z^2+T_x^2)\Si_\Omega^2\geq \Si_\Omega^2 T_z^2 + \C_\Omega^2 T_x^2 \geq \beta^2$. Thus the derivative is positive and the length of ${\bf v}_\gamma$ is increasing with $\gamma$.  

We can now give a simple geometrical interpretation to our problem of finding the minimal $\phi$: for each value $\phi$ such that 
\be
\binom{\C_\phi}{\Si_\phi}^T {\bf v}_\gamma \geq 1
\ee
a line tangent to the unit circle at $\binom{\C_\phi}{\Si_\phi}$ is also crossing the curve ${\bf v}_\gamma$. So for the minimal value $\phi_*$ there is a line tangent to both ${\bf v}_\gamma$ and the unit circle at $\binom{\C_{\phi_*}}{\Si_{\phi_*}}$. The equation of the line tangent to ${\bf v}_\gamma$ reads ${\bm \ell}(\lambda)= {\bf v}_\gamma + \lambda {\bf v}_\gamma'$. This line is tangent to the unit circle if and only if the equation 
\be
|{\bf v}_\gamma + \lambda {\bf v}_\gamma'|^2 = 1
\ee
has only one solution, where the derivative is with respect to $\gamma$. This is a quadratic equation
\be
\lambda^2 |{\bf v}_\gamma'|^2 + 2 \lambda {\bf v}_\gamma' \cdot {\bf v}_\gamma +  |{\bf v}_\gamma|^2 - 1 =0,
\ee
which has a unique solution iff its determinant is zero
\be
\left({\bf v}_\gamma' \cdot {\bf v}_\gamma\right)^2 = |{\bf v}_\gamma'|^2 \left(|{\bf v}_\gamma|^2 - 1 \right).
\ee
With 
\be
{\bf v}_\gamma'=\frac{\C_\Omega}{(\beta^2-\Si_\Omega^2(T_x^2 \C_\gamma^2 + T_z^2 \Si_\gamma^2))^{3/2}}\binom{T_z \Si_\gamma(T_z^2 \Si_\Omega^2-\beta^2)}{- T_x \C_\gamma(T_x^2 \Si_\Omega^2 -\beta^2)}
\ee
lengthy but straightforward algebra gives
\be
\Si_{\gamma}^2 = -\frac{T_x^2(\beta^2 - T_x^2 \Si_\Omega^2)(\beta^2 - (T_z^2+T_x^2)\Si_\Omega^2 )}{(T_z^2-T_x^2)\left(\beta^4 + \Si_\Omega^2(T_x^2 T_z^2 - 2\beta^2(T_x^2+T_z^2))+ \Si_\Omega^4 (T_x^4 + T_z^4)\right)}.
\ee
 To find that the minimal angle $\phi_*$ note that for the tangent line $\binom{\C_{\phi_*}}{\Si_{\phi_*}}\cdot {\bf v}_\gamma'=0$, and therefore $\binom{-\Si_{\phi_*}}{\C_{\phi_*}}=\frac{{\bf v}_\gamma'}{|{\bf v}_\gamma'|}$. Plugging in the above equations we find
\be
c_*^2({\bf L}, \Omega,\beta)= \cos^2\big(\phi_*({\bf L}, \Omega,\beta)\big)= \frac{(\beta^2-\Si_\Omega^2 T_x^2)(\C_\Omega^2 T_x^2 + \Si_\Omega^2 T_z^2-\beta^2)}{\C_\Omega^2 (T_z^2-T_x^2)(\Si_\Omega^2 T_z^2 + \Si_\Omega^2 T_x^2 - \beta^2)}.
\ee

\subsection{Eve's maximum information for $\Omega\leq \frac{\pi}{4}$}
\label{app: analytic entropy}
We here give details on the derivation of the formula~\eqref{easy_sol} in the main text which corresponds to Eve's maximum information in the case $\Omega\leq \frac{\pi}{4}.$ For these inequalities we have seen that the constraint  $\C_\Omega^2\,  T_z^2 +\Si_\Omega\,  T_x^2\geq \beta^2$ can be fulfilled with $c^2_*({\bf L},\beta) =1$. It has been shown in Ref.~\cite{Ho20} that $H(\rho_{E|\hat{a}=+1})$ is a monotonic function in the key generating setting $\phi \in [0,\frac{\pi}{4}]$. The wost case (optimal attack for Eve) thus consists in setting the measurement angle $\C_\phi=1$, which implies a simple form for the state
\be
\rho_{E|\hat{a}=+1}= \left(
\begin{array}{cccc}
 L_1 & 0 & \sqrt{L_1 L_3 q} & 0 \\
 0 & L_2 & 0 & -\sqrt{L_2 L_4 q} \\
 \sqrt{L_1 L_3 q} & 0 & L_3 & 0 \\
 0 & -\sqrt{L_2 L_4 q} & 0 & L_4 \\
\end{array}
\right),
\ee
with a closed form expression for its eigenvalues implying
\be
H(\rho_{E|\hat{a}=1}) = H
\Big({\bf p} = \left(
\begin{array}{c}
     \frac{1}{2} \left(L_1+L_3+\sqrt{4 L_1 L_3 q+\left(L_1-L_3\right){}^2}\right)\\
    \frac{1}{2} \left(L_1+L_3-\sqrt{4 L_1 L_3 q+\left(L_1-L_3\right){}^2}\right) \\
     \frac{1}{2} \left(L_2+L_4+\sqrt{4 L_2 L_4 q+\left(L_2-L_4\right){}^2}\right)\\
     \frac{1}{2} \left(L_2+L_4-\sqrt{4 L_2 L_4 q+\left(L_2-L_4\right){}^2}\right)
\end{array}\right)\Big).
\ee
The constraint on the generalized CHSH score leads to the following constraint on the vector $\bf L$ 
\be
\binom{T_Z^2}{T_X^2} \cdot \binom{\C_\Omega^2}{\Si_\Omega^2} = 
2 \left(L_3-L_4\right) \left(L_1-L_2\right)\C_{2\Omega}+\left(L_1-L_2\right){}^2+\left(L_3-L_4\right){}^2 \geq \beta^2.
\ee
Our goal is thus to find the components of the vector $\bf L$ maximizing $H(\bf L) - H(\bf p)$ and satisfying the previous constraint. Inspired by Ref.~\cite{Ho20}, we first introduce the following parametrization
\be
\begin{split}
L_1 &= P x\\
L_3 &= P(1-x)\\
L_2 &= (1-P)y\\
L_4 &= (1-P)(1-y).
\end{split}
\ee
The partial ordering of the $\bf L$ coefficients implies
\be
(1-P)y \leq P x \leq (1-P)y + 2P-1
\ee
which requires $P\geq \frac{1}{2}.$ The advantage of this parametrization comes from the fact that our figure of merit can be nicely rewritten as
\begin{equation}
   \begin{split}\label{eq: def hq}
H({\bf L}) - H({\bf p}) &= P h_q(x) + (1-P) h_q (y) \\
 h_q(z) & = h(z) - h(n_q(z)) \\
 n_q(z) &=\frac{1 + \sqrt{1 - 4\, (1-q)\, z (1-z)}}{2}\\
  \end{split}
\end{equation}
where $ h(z) = -z \log(z) - (1-z) \log(1-z)  $ is the binary entropy function with the logarithm in base 2, while the constraint on the expected value of the generalized CHSH operator is given by
\be
\Si_\Omega\,  (2 P (x+y-1)-2 y+1)^2+ \C_\Omega \, (1-2 P)^2 \geq \beta^2
\ee
For a fixed $P,$ the curve in the $(x,y)$-plan that corresponds to a constant value $\beta$ satisfies $P \, dx = (1-P)\, dy$. This remark allows one to maximize $H({\bf L}) - H({\bf p})$  along this curve and find that it is optimal for Eve to set $x+y =1$, see appendix C2 in Ref.~\cite{Ho20} for the detailed argument. The symmetry of the function $h_q(x)=h_q(1-x)$ allows us to write the problem as  
\be\begin{split}
\max_{x,P}& \ \, H({\bf L}) - H({\bf p}) = h_q(x)\\
\text{s.t.}& \ \, (2P-1)^2 \C_\Omega^2+(1-2 x)^2 \Si_\Omega^2 \geq \beta^2.
\end{split}
\ee
As the goal function $h_q(x)$ does not depend on $P$, we can set its value to $P=1$ because this is the value maximizing the LHS of the constraint inequality and allowing the largest possible interval for the remaining variable $x$. We thus get
\be\begin{split}
\max_{x,P}& \ \, H({\bf L}) - H({\bf p}) = h_q(x)\\
\text{s.t.}& \ \, (1-2 x)^2  \geq  \frac{\beta^2 -\C_\Omega^2}{\Si_\Omega^2}.
\end{split}
\ee
Finally, as $h_q(x)$ is a monotonically decreasing function of $x$ (see Ref.~\cite{Ho20}), it is optimal to set $x$ to the least possible value compatible with the constraint. This implies 
\be
\boxed{
\begin{split}
I(\beta;\Omega, q) &= h_q(z) \\
\text{with} \ \,
z &= \frac{1}{2} \left(\frac{\sqrt{ \beta ^2-\C_{\Omega}^2}}{ \Si_\Omega }+1\right).
\end{split}}
\ee\\

Let us now recall that the situation with $c^2_*({\bf L},\beta) =1$ and $\C_\Omega^2\,  T_z^2 +\Si_\Omega\,  T_x^2\geq \beta^2$ also occurs in the case $\Omega>\frac{\pi}{4}$. The above proof guarantees that it is optimal for Eve to use strategies where the inequality is saturated $\C_\Omega^2\,  T_z^2 +\Si_\Omega\,  T_x^2 = \beta^2$. Hence in the optimization problem for $\Omega>\frac{\pi}{4}$ we can ignore all the strategies with $\C_\Omega^2\,  T_z^2 +\Si_\Omega\,  T_x^2 > \beta^2.$



\subsection{Concavity of $h_q \circ z(\beta)$}

\label{app: convex}

Recall that in order to use the bound $I(\beta;\Omega,q)$ derived for two-qubit strategies in the previous section as a universal bound (valid for strategies in any dimension), we have to show that this function is concave. In this case, for any mixture of qubit strategies (enforced by the Jordan's lemma) with an average score $\bar \beta= \sum_i p_i \beta_i$, Eve's information satisfies
\be
\bar I(\beta;\Omega,q) = \sum_i p_i I(\beta_i;\Omega,q)\leq I(\bar \beta;\Omega,q).
\ee
The concavity of $I(\beta;\Omega,q) =h_q(z(\beta))$ follows from the fact that its second derivative is negative
\be
\frac{d^2}{d\beta^2} h_q\big(z(\beta)\big) = h_q''(z) \big(z'(\beta)\big)^2 + h_q'(z)z''(\beta)\leq 0,
\ee
which we are going to show below.
In this section we will use the natural algorithm instead of logarithm in base 2.  The function $h_q(z)$ takes a positive real factor upon changing the base of the algorithm, so it is irrelevant for its concavity. Note first that $z(\beta)\in[\frac{1}{2},1]$ and 
\be
\sqrt{\beta^2-\C_\Omega^2} = \Si_\Omega(2z-1).
\ee
Then consider the following identities
\be
\begin{split}
    (z'(\beta))^2 &= \frac{\beta^2}{4 \Si_\Omega^2 (\beta^2-\C_\Omega^2)} = \frac{\Si_\Omega^2(2z-1)^2+\C_\Omega^2}{4 \Si_\Omega^4(2z-1)^2}\\
    z''(\beta) &= \frac{-\C_\Omega^2}{2\Si_\Omega (\beta^2-\C_\Omega^2)^{3/2}}= \frac{-\C_\Omega^2}{2\Si_\Omega^4 (2z-1)^3}.
    \end{split}
\ee
The identity~\eqref{eq: convex proof} that we want to prove thus becomes
\be
\frac{h_q''(z)(2z-1)(\Si_\Omega^2 (2z-1)^2+\C_\Omega^2)- 2 h_q'(z) \C_\Omega^2}{4 \Si_\Omega^4 (2z-1)^3} \leq 0.
\ee
Multiplying by a positive fraction $\frac{4 \Si_\Omega^4 (2z-1)^3}{\C_\Omega^2}$ it can be straightforwardly simplified  to the form 
\be\label{eq: cond1}
h_q''(z)(2z-1)(\text{T}_\Omega^2 (2z-1)^2+1)- 2\,  h_q'(z) \leq 0.
\ee
As  $h_q''(z)\leq 0$ was proven in~\cite{Ho20,Woodhead14}, we have the inequality 
\be
h_q''(z)(2z-1)^2\, \text{T}_\Omega^2 \leq 0.
\ee
We use it to relax the inequality in Eq.~\eqref{eq: cond1} to
\be
h_q''(z)(z-\frac{1}{2}) -  h_q'(z) \leq 0.
\ee
At the point $z= \frac{1}{2}$ the left hand side becomes zero, since $h_q'(\frac{1}{2})=0$ and $|h_q''(\frac{1}{2})|< \infty,$ see below.
From now on, we thus exclude the point $z=\frac{1}{2}$ and consider $z \in (\frac{1}{2},1]$. Now we can divide the whole expression by a strictly positive  factor $(z-\frac{1}{2})^2$. We obtain
\be
\frac{1}{2} \frac{h_q''(z)(z-\frac{1}{2}) -  h_q'(z)(z-\frac{1}{2})'}{(z-\frac{1}{2})^2}\leq 0,
\ee
or
\be
\frac{1}{2}\left(\frac{h_q'(z)}{z-\frac{1}{2}}\right)'=\left(\frac{h_q'(z)}{2z-1}\right)'\leq 0.
\ee
In other words we want to show that the function
\be
(\ast) \, \frac{d}{dz} \left(\frac{h_q'(z)}{2z-1}\right) \leq 0
\ee
on the interval $z\in(\frac{1}{2},1]$. Let us now compute this function:
\be
\frac{h_q'(z)}{2z-1} = \frac{h'(z)}{2z-1} - h'(n_q(z)) \frac{n_q'(z)}{2z-1}.
\ee
The last fraction can be simplified to
\be\label{eq:justBefore}
    \frac{n_q'(z)}{2 z-1} = \frac{(2z-1)(1-q)}{(2 z-1)\sqrt{1 - 4\, (1-q)\, z (1-z)}} = \frac{1-q}{2n_q(z)-1}.
\ee
Therefore
\be
    \frac{h_q'(z)}{2z-1} =\frac{h'(z)}{2z-1} - (1-q) \frac{h'(n_q(z))}{2 n_q(z)-1} = g(z) - (1-q) g\big(n_q(z)\big),
\ee
where
\be
g(z) = \frac{h'(z)}{2z-1}= - \frac{\log(\frac{z}{1-z})}{2z-1}.
\ee 

To complete the proof we thus need to show that
\be
(\ast)\quad g'(z) - (1-q)g'\big( n_q(z)\big) n_q'(z)\leq 0.
\ee
From Eq.~\eqref{eq:justBefore}, we have
\be\label{eq: nq prime}
n_q'(z)= (1-q)\frac{2z-1}{2 n_q(z)-1},
\ee
so the inequality to be shown can be rewritten as
\be
\frac{g'(z)}{2z-1} - (1-q)^2\frac{g'(n_q(z))}{2  n_q(z)-1}\leq 0.
\ee
For $q=0$ we have $(1-q)=1$ and $n_q(z)=z$ so the two terms are equal. The identity to be shown can then be expressed as
\be\begin{split}
(\ast)\qquad\ \ \  f(z,0)& - f(z,q)\leq 0\\
\text{with} \quad f(z,q)&= (1-q)^2 u(n_q(z))\\
u(z)&= \frac{g'(z)}{2z-1}=\frac{\frac{1}{z-1}+\frac{1}{z}+2 \log \left(\frac{z}{1-z}\right)}{(2 z-1)^3},
\end{split}
\ee
which holds for $q=0$ trivially. To show that it holds for all $q$ it is sufficient to demonstrate that the function $f(z,q)$ is increasing with $q$, i.e.
\be
(*)\qquad \frac{d}{dq} f(z,q)\geq 0,
\ee
which we are going to show now. \\
Using 
\be\begin{split}
\frac{d}{dq} n_q(z) = \frac{(1-z) z}{\sqrt{1-4 (1-q) (1-z) z}} 
=\frac{n_q(z)(1-n_q(z))}{(1-q)(2n_q(z)-1)}
\end{split}\ee
we obtain 
\be\begin{split}
\frac{d}{dq} f(z,q) &= - 2 (1-q) u\big(n_q(z)\big)+ (1-q)^2 u'\big(n_q(z)\big)\frac{d}{dq} n_q(z)\\
& = - 2(1-q)u(n) + (1-q) u'(n) \frac{n(1-n)}{2n -1}
\end{split}
\ee
for $n > \frac{1}{2}$, which is positive iff $q=1$ or
\be
(\ast)\qquad u'(n) n(1-n) - 2 u(n) (2n-1) \geq 0.
\ee
In the case $q<1$, straightforward algebra allows to find a simple expression of the left hand side and rewrite the condition as
\be
\begin{split}
&\frac{6 n^2 - 4 n^3-1 + 4 (n^4-2 n^3+2 n^2-n) \log \left(\frac{n}{1-n}\right)}{(2n-1)^4 n (1-n)}\geq 0 \\
\iff \ \ & 6 n^2 - 4 n^3-1 + 4 (n^4-2 n^3+2 n^2-n) \log\left(\frac{n}{1-n}\right) \geq 0
\end{split}
\ee
 Changing the variable to $x+1=\frac{n}{1-n}$ with $x\geq 0$ we express the above inequality as
\be
\begin{split}
&\frac{x (x^2+6x +6)}{(x+2)^3} - 4 \frac{(x+1) (x^2+3x+3)}{(x+2)^4}\log (x+1) \geq 0\\
\iff \ \ &\log(1+x)\leq \frac{x (x+2) (x^2+6 x+6)}{4 (x+1) (x^2+3 x+3)}.
\end{split}
\ee
To prove this relation note that there is equality for $x=0$, but the LHS increases slower than the RHS
\be\begin{split}
&\log'(1+x)\leq \left(\frac{x (x+2) (x^2+6 x+6)}{4 (x+1) (x^2+3 x+3)}\right)'\\
\iff \ \ & \frac{1}{1+x}\leq \frac{1}{4} \left(\frac{9 (x+1)}{(x (x+3)+3)^2}+\frac{1}{(x+1)^2} +\frac{3}{x (x+3)+3}+1\right)\\
\iff \ \ & \frac{x^4 (x+2)^2}{4 (x+1)^2 (x (x+3)+3)^2} \geq 0,
\end{split}
\ee
which concludes the proof.\\

\paragraph{Properties of $h_q'(\frac{1}{2})$ and $h_q''(\frac{1}{2})$} 
We start with the first derivative and want to show that $h_q'(\frac{1}{2})=1/2.$ We have
\be
h_q'(z) = \Big( h(z)- h(n_q(z))\Big)' = h'(z) - h'(n_q(z)) n_q'(z)
\ee
The binary entropy hits a maximum at $z=1/2$ so $h'(\frac{1}{2})=0$. For the second term, we use \eqref{eq: nq prime} to get 
\be
h'(n_q(z)) n_q'(z) = - (1-q)\frac{2z-1}{2 n_q(z)-1} \log\left(\frac{n_q(z)}{1-n_q(z)}\right ),
\ee
$(2z-1)=0$ at $z=\frac{1}{2}$, while  
\be
\left|\left.\frac{1-q}{2 n_q(z)-1} \log\left(\frac{n_q(z)}{1-n_q(z)}\right ) \right|_{z=1/2}\right| = \left|\frac{(1-q) \log \left(\frac{1+\sqrt{q}}{1-\sqrt{q}}\right)}{\sqrt{q}}\right|.
\ee
Changing the variable to $x+1=\frac{1+\sqrt{q}}{1-\sqrt{q}}$ with $x\geq 0$ yields for the last expression
\be
\left|\frac{4 (x+1) \log (x+1)}{x (x+2)}\right| < \infty.
\ee
It is obviously bounded for $x>\epsilon$ with any $\epsilon$, and the fact that  the limit $x\to 0$ exists can be seen by straightforward application of l'H\^{o}pital's rule. Hence,
\be
h_q'\left(\frac{1}{2}\right)=0.
\ee\\

We also wish to show that the second derivative $h_q''(\frac{1}{2})$ is bounded. To do so we compute
\be
h_q''\left(\frac{1}{2}\right) = \frac{2 (1-q) \log \left(\frac{1+\sqrt{q}}{1-\sqrt{q}}\right)}{\sqrt{q}} - 4.
\ee
But as we have just shown that 
\be
\left|\frac{ (1-q) \log \left(\frac{1+\sqrt{q}}{1-\sqrt{q}}\right)}{\sqrt{q}}\right| <\infty,
\ee
the desired result
\be
\left|h_q''\left(\frac{1}{2}\right)\right|<\infty
\ee
follows.

\subsection{Maximization of the generalized CHSH score with respect to auxiliary settings}
 \label{app: best score}
In the goal function 
\be
H(\rho_E) - H(\rho_E|\hat{a}_0(q)) + t\,  \mathcal{B}_\Omega({\bf L},\phi,{\bm a}_1, {\bm b}_0, {\bm b}_1)
\ee
that \textit{a priori} appears in Eq.~\eqref{eq: dual qubit}, it is only the Bell score which depends on the auxiliaty measuremnt settigns ${\bm a}_1, {\bm b}_0$ and ${\bm b}_1$.  As $t$ is always positive we can straigtforwardly maximise the score with respect to these settings. We thus define
\be
\beta_\text{max}({\bm L},\phi) = \max_{{\bm a}_1, {\bm b}_0, {\bm b}_1} \mathcal{B}_\Omega({\bf L},\phi,{\bm a}_1, {\bm b}_0, {\bm b}_1),
\ee
which actually appears in Eq.~\eqref{eq: dual qubit}. 

Let us now compute this expression starting from Eq.~\eqref{eq: Bell score full}, that we put in the form
\be
\mathcal{B}_\Omega  =  \binom{\C_\theta}{ \Si_\theta} \cdot
\left(\begin{array}{c}
\C_\Omega\,  {\bf a}_0^T \binom{T_z \C_\gamma}{ T_x \Si_\gamma}
\\
 \Si_\Omega \, {\bf a}_1^T
\binom{-T_z \Si_\gamma}{T_x \C_\gamma}
\end{array}\right).
\ee
This form makes the maximization with respect to $\theta$ and ${\bf a}_1$ straightforward
\be\begin{split}
    \max_{\theta,{\bf a_1}} \mathcal{B}_\Omega  &=\max_{{\bf a}_1}\sqrt{\left(\C_\Omega\,  {\bf a}_0^T \binom{T_z \C_\gamma}{ T_x \Si_\gamma}\right)^2 + \left( \Si_\Omega \, {\bf a}_1^T
\binom{-T_z \Si_\gamma}{T_x \C_\gamma} \right)^2 }
\\
& =\sqrt{\C_\Omega^2 \left( \C_\phi T_z \C_\gamma + \Si_\phi T_x \Si_\gamma\right)^2 +  \Si_\Omega^2 
\left( T_z^2 \Si_\gamma^2 + T_x^2 \C_\gamma^2 \right) }.
\end{split}
\ee
To find the maximum with respect to $\gamma$ or ${\bf c}=\binom{\C_\gamma}{\Si_\gamma}$ it is convenient to write the expression inside the square root as 
\be
\left(\max_{\theta,{\bf a_1}} \mathcal{B}_\Omega \right)^2 = {\bf c}^T
\left(\begin{array}{cc}
\C_\Omega^2 \C_\phi^2 T_z^2 + \Si_\Omega^2 T_x^2 & \C_\Omega^2 \C_\phi \Si_\phi T_z T_x\\
\C_\Omega^2 \C_\phi \Si_\phi T_z T_x & \C_\Omega^2 \Si_\phi^2 T_x^2 + \Si_\Omega^2 T_z^2
\end{array}\right)
{\bf c}.
\ee
It is now obvious that the value is maximal if ${\bf c}$ is aligned with the eigenvector of the matrix which corresponds to the maximal eigenvalue. Therefore
\be\small\begin{split}
\beta_\text{max}({\bm L},\phi) &= \max_{\theta,\gamma,{\bf a}_1} \mathcal{B}_\Omega({\bf L},\phi,{\bm a}_1, {\bm b}_0, {\bm b}_1),\\ &=\sqrt{\text{Eig}_+\left(\begin{array}{cc}
\C_\Omega^2 \C_\phi^2 T_Z^2 + \Si_\Omega^2 T_x^2 & \C_\Omega^2 \C_\phi \Si_\phi T_z T_x\\
\C_\Omega^2 \C_\phi \Si_\phi T_z T_x & \C_\Omega^2 \Si_\phi^2 T_x^2 + \Si_\Omega^2 T_z^2
\end{array}\right)}\\
&=\frac{1}{\sqrt{2}}\left(\C_\Omega^2 (\C_\phi^2 T_z^2 +\Si_\phi^2 T_x^2) +\Si^2_\Omega (T_z^2+T_x^2)+\sqrt{(\C_\Omega^2 (\C_\phi^2 T_z^2 -\Si_\phi^2 T_x^2) -\Si^2_\Omega (T_z^2-T_x^2))^2+4 (\C_\Omega^2\C_\phi \Si_\phi T_z T_x )^2 }\right)^{1/2}.
\end{split}\ee\\

\section{Numerical tool}

\subsection{Lipshitz continuity of von Neumann entropy}

\label{app: Lipsitz entropy}
Consider two states $\rho$ and $\sigma$ on an $n$-dimensional Hilbert space, that are close in  fidelity:
\be
F(\rho,\sigma) = \text{tr} \left|\sqrt{\rho} \sqrt{\sigma}\right| = f.
\ee
Given the monotonicity of $\arccos$ in the range $[0,1]$, this condition can be equivalently written in terms of the angle $A(\rho,\sigma) = \arccos(F(\rho,\sigma))$
\be
A(\rho,\sigma) = a =\arccos(f).
\ee
The angle is a metric on the space of density operators~\cite{Nielsen02}, in particular it satisfies the triangle inequality.\\

Next, note that the angle between two states is lower bounded by the angle between 
the ordered vectors made of their ordered eigenvalues
\be
a  =A(\rho, \sigma) \geq A({\bf p }, {\bf q})= \arccos\left({\sqrt{\bf p} \cdot \sqrt{\bf q}}\right),
\ee
with ${\bf p} = \text{Eig}^\downarrow(\rho)$, ${\bf q} = \text{Eig}^\downarrow(\sigma)$ such that $p_1\geq p_2\geq\dots$ This inequality follows from 
\be
    F(\rho,\sigma) =\text{tr} |\sqrt{\rho} \sqrt{\sigma}| = \max_U \text{tr} \sqrt{\rho} \, (\sqrt{\sigma} U) \leq \sqrt{\bf p} \cdot \sqrt{\bf q},
\ee
where in second line we used the so-called von Neumann trace inequality~\cite{Mirsky1975}. This bound is useful because the entropies of the states match the entropies of the probability distributions
\be
H(\rho) = H({\bf p})\ , \qquad H(\sigma) = H({\bf q}).
\ee
Let us now bound their difference
\be
\Delta H = \left|H(\rho)- H(\sigma)\right| = \left|H({\bf p}) -  H({\bf q})\right|.
\ee
To do so, note that for any two unit vectors $\sqrt{\bf p}$ and $\sqrt{\bf q}$ on the $n$-sphere there exist a path $\bm \gamma$ connecting the two and such that the integral along the path satisfies
\be
\int_{\sqrt{\bf p}}^{\sqrt{\bf q}} dA =  A({\bf p }, {\bf q}) \leq a.
\ee

Let us bound the variation of the entropy along the path. To this end, we associate a probability distribution ${\bf r}$ to each vector ${\bf v}$ on the path $\bm \gamma$, with $r_i=(v^{(i)})^2$ (note that the vectors along the curve remain in the positive part of the $n$-sphere $v^{(i)}\geq 0$). A step $dA$ from $\bf v$ along the path corresponds to some deformation of the vector given by 
\be
{\bf v}_{dA} \to {\bf v} + {\bf v}_\perp dA,
\ee
with ${\bf v}\cdot {\bf v}_\perp =0$. To simplify the following computations we introduce the ``natural'' entropy 
\be
H_e(\rho)= - \t{tr} \rho \ln(\rho)= \ln(2) \, H(\rho),
\ee
as the von Neumann entropy computed with the natural logarithm (recall that the $\log$ in $H(\rho)$ was taken in base 2).
From
\be
H_e({\bf r}) = - \sum_i (v^{(i)})^2 \ln((v^{(i)})^2) 
\ee
we compute the entropy variation for an infinitesimal increment of the angle
\be\begin{split}
\left|\frac{dH_e}{dA}\right| &= 2 |\sum_i v^{(i)} v_\perp^{(i)}( \ln( (v^{(i)})^2) +1)| \\
&=2 |\sum_i v_\perp^{(i)} \, v^{(i)} \ln( (v^{(i)})^2)| \\
&= 2 \, |{\bf v}_\perp \cdot {\bf w}| \leq 2|| {\bf w}|| 
\end{split}
\ee
where we defined a vector $\bf w$  as $w_i =  v^{(i)} \ln((v^{(i)})^2) = \sqrt{r_i} \ln(r_i)$. Hence we obtain
\be
\left|\frac{dH_e}{dA}\right| \leq 2\sqrt{ \sum_i r_i \ln^2(r_i)},
\ee
and it remains to bound the expression on the RHS. To do so, we will construct a concave upper bound on the function
\be
c(r) =  r\ln^2(r)
\ee
defined on $[0,1]$. By computing the second derivative
\be
c''(r) = 2 \frac{1+\ln(r)}{r}
\ee
we see that the function is concave $c''\leq 0$ on the interval $r\in[0, \frac{1}{e}]$, and convex $c''\geq 0$ on the complement. To get a concave upper-bound we thus look for a line passing by $r=1$ and $c(1)=0$ and tangent to $c(r)$. In the $(r,c)$-plane the equation of a line tangent to $c(r)$ at $r$ is given by
\be
{\bm \ell}_r(\lambda)=\binom{r}{c(r)}+ \binom{1}{c'(r)} \lambda.
\ee
It passes through the point ${\bm \ell}_r(\lambda)= (1,0)$ iff
\begin{align}
&\begin{cases}
r +\lambda = 1 \\
c(r) + \lambda c'(r) = 0
\end{cases} \nonumber\\
\implies & c(r) + (1-r) c'(r) =0 \nonumber\\
\implies & \ln(r) (2 -2 r+\ln(r)) =0 \nonumber\\
\implies &\begin{cases}
r_1 = -\frac{1}{2} W_0(-\frac{2}{e^2})\approx 0.203\\
r_2= 1
\end{cases}, \label{eq:r1value}
\end{align}
where $W_0$ is the principal branch of the Lambert $W$-function, and the trivial solution $r_2=1$ is irrelevant. Knowing $r_1$ we can construct a concave upper bound
\be
c(r)\leq \hat{c}(r) =
\begin{cases}
c(r) & r\leq r_1\\
c(r_1)\frac{1-r}{1-r_1} & r> r_1.
\end{cases}
\ee
Note that $r_1< \frac{1}{e}$. Furthermore, $r_1$ is the solution of the equation $2-2r_1 +\ln(r_1)=0$. This allows us to simplify
\be
c(r_1) = r_1 \ln^2(r_1)= 4 \, r_1 (1- r_1)^2,
\ee
and
\be\label{eq: c func}
\hat{c}(r) =
\begin{cases}
r \ln^2(r) & r\leq r_1\\
4\,  r_1 (1-r_1)(1-r) & r> r_1.
\end{cases}
\ee\\

Finally with the concave bound $\hat{c}$ it is easy to obtain
\be\begin{split}
\sum_i r_i \ln^2(r_i) &= \sum_i c(r_i)\\
&\leq \sum_i \hat{c}(r_i) =n \sum_i \frac{1}{n} \hat{c}(r_i)\\
& \leq n\,  \hat{c}\left( \sum_i \frac{r_i}{n}\right) = n\,  \hat{c}\left( \frac{1}{n}\right).
\end{split}
\ee
So for entropy susceptibility we get 
\be
\left|\frac{dH_e}{dA}\right|\leq 2 \sqrt{n \, \hat{c}\left(\frac{1}{n}\right)}
\ee
For $n\leq 4$, we have $r_1 < \frac{1}{n}$, so from Eq.~\eqref{eq: c func} we get 
\be
n\leq 4: \qquad n \, \hat c\left(\frac{1}{n}\right) = 4\,  r_1(1-r_1)(n-1).
\ee
For $n\geq 5$, we instead have $r_1 > \frac{1}{n}$, so we are in the other ``part'' of the function in Eq.~\eqref{eq: c func} and get a simpler expression:
\be
n\geq5 :\qquad n\,\hat c\left(\frac{1}{n}\right) =\ln^2(n).
\ee
Finally, combining the two and recalling that $H(\rho) = \frac{H_e(\rho)}{\ln(2)}$ we get the expression
\be\boxed{
\left|\frac{dH}{dA}\right|\leq 
\begin{cases}
\frac{4\sqrt{r_1(1-r_1)}}{\ln(2)} \sqrt{n-1} & n\leq 4 \\
2 \log(n) &  n \geq 5
\end{cases}
}.
\label{eq:gradbound}
\ee
For later use, we evaluate the constant for the $n=4$ case in particular:
\be
\left|\frac{dH}{dA}\right| \leq \frac{4\sqrt{r_1(1-r_1)}}{\ln(2)} \sqrt{3} < 4.023.
\ee

To bound the entropy difference for non-infinitesimal distances, we simply integrate along the curve $\bm \gamma$; e.g.~for $n=4$ this yields
\be
\Delta H 
\leq \int_{\sqrt{\bf p}}^{\sqrt{\bf q}} \left|\frac{dH}{dA}\right|\,  dA 
\leq 4.023\,  A(\rho,\sigma).
\ee

\subsection{Continuity of the goal function}

\label{app: continuity of goal}
\paragraph{The entropy term --}

To apply the continuity bound previously described to our situation, let $\rho,\rho'$ be the states on $\hat{A}_0 E$ produced by measurements along angles $\phi,\phi'$ on the states $\ket{\Psi}_{ABE}$ with the weight ${\bf L}, {\bf L}'$ respectively. Our aim is to bound $\left|H(\hat{A}_0| E)_{\rho} - H(\hat{A}_0| E)_{\rho'}\right|.$ We use
\be\begin{split}
    \left|H(\hat{A}_0| E)_{\rho} - H(\hat{A}_0| E)_{\rho'}\right| &=
    \left|H(\rho_E) - H( E|\hat A_0)_{\rho } - H(\rho'_E) + H( E|\hat A_0)_{\rho '} \right|
    \\ &\leq \left|H(\rho_E)  - H(\rho'_E) \right| +
    \left| H( E|\hat A_0)_{\rho } - H( E|\hat A_0)_{\rho '} \right|
\end{split}
\ee
and bound the two last terms independently. For the first term involving $H(\rho_E)_\rho$, things are straightforward. The states of Eve $\rho_E = \text{diag}({\bf L})$ are 4-dimensional and independent of $\phi$, so 
\be
|H(\rho_E)_{\bf L} + H( \rho_E)_{\bf L '} |\leq 4.023 \arccos(\sqrt{\bf L}\cdot \sqrt{ \bf L '}).
\ee

For the other term, we note
\be\begin{split}
\left| H( E|\hat A_0)_{\rho } - H( E|\hat A_0)_{\rho '} \right| &= \left| H(\hat A_0 E)_{\rho } - H(\hat A_0 E)_{\rho '} \right|\\
&= \left| H( \rho ) - H(\rho ') \right|
\end{split}
\ee
as $H(\hat A_0)=1$. Since $\hat A_0 E$ is 8-dimensional we get
\be
    \left| H( E|\hat A_0)_{\rho } - H( E|\hat A_0)_{\rho '} \right| \leq 2 \log(8) \, A(\rho, \rho') =6\, A(\rho, \rho').
\ee
Given that $\rho=\rho({\bf L},\phi)$ and $\rho'= \rho({\bf L}',\phi')$, we can use the triangle inequality to write
\be
A(\rho, \rho') \leq A(\rho({\bf L},\phi),\rho({\bf L}',\phi))+
 A(\rho({\bf L}',\phi),\rho({\bf L}',\phi')).
\ee
For $A(\rho({\bf L},\phi),\rho({\bf L}',\phi))$ we note that both states result from the action of the same CPTP map on two initial state $\ket{\Psi({\bf L})}_{ABE}$ and $\ket{\Psi({\bf L}')}_{ABE}$. By the data-processing inequality (inherited by the angle from the fidelity $F$) we have 
\be\begin{split}
     A(\rho({\bf L},\phi),\rho({\bf L}',\phi)) & \leq A(\ket{\Psi({\bf L})}_{ABE}, \ket{\Psi({\bf L}')}_{ABE})\\
     & = A({\bf L},{\bf L'})\\
     & = \arccos(\sqrt{\bf L}\cdot \sqrt{ \bf L '}).
\end{split}
\ee
To bound the other term $A(\rho({\bf L}',\phi),\rho({\bf L}',\phi'))$ we note that if we apply a channel that performs a $Z$ measurement followed by noisy preprocessing to the state $\left(e^{i\phi Y_A/2} \otimes \Id_{BE}\right)\ket{\Psi}_{ABE}$, this produces exactly the state $\rho$ on $\hat{A}_0 E$; analogously, applying the same channel to the state $\left(e^{i\phi' Y_A/2} \otimes \Id_{BE}\right)\ket{\Psi}_{ABE}$ produces the state $\rho'$.
Therefore the data-processing inequality implies $F\left(\rho,\rho'\right)$ is lower-bounded by
\be\begin{split}
F \left(\left(e^{i\phi Y_A/2} \otimes \Id_{BE}\right)\ket{\Psi}, \left(e^{i\phi' Y_A/2} \otimes \Id_{BE}\right)\ket{\Psi}\right) =& \left|\bra{\Psi} \left(e^{i (\phi'-\phi) Y_A/2} \otimes \Id_{BE}\right) \ket{\Psi}\right| \\
=& \left|\cos\frac{\phi'-\phi}{2} + i \sin\frac{\phi'-\phi}{2} \bra{\Psi} \left(Y_A \otimes \Id_{BE}\right) \ket{\Psi}\right|\\
\geq& \left|\cos\frac{\phi'-\phi}{2}\right|, \quad \text{since 
$\bra{\Psi} \left(Y_A \otimes \Id_{BE}\right) \ket{\Psi} \in \mathbb{R}$.
}
\end{split}\ee
Putting these together, we conclude that (for $|\phi'-\phi|<\pi$)
\begin{align}
A(\rho({\bf L}',\phi),\rho({\bf L}',\phi')) \leq
\frac{|\phi'-\phi|}{2}.
\end{align}
Combining everything together, we get 
\be\label{eq: bound cont H}
    \left|H(\hat{A}_0| E)_{\rho} - H(\hat{A}_0| E)_{\rho'}\right| \leq 10.023\,  \arccos(\sqrt{\bf L}\cdot \sqrt{ \bf L '}) + 3\,  |\phi-\phi'|
\ee\\

\paragraph{The Bell score -- }

Next we wish to bound the increment $|d \beta_\text{max}({\bf L, a_0},\Omega)|$ for infinitesimal changes of the parameters $(d{\bf L}, d\phi)$. It is actually straightforward to bound the gradient of the Bell score before the maximization with respect to  ${\bf a}_0,{\bf b}_0, {\bf b}_1$, so we just need to be careful to apply this bound on $\beta_\text{max}$.

First, note that $\mathcal{B}_\Omega({\bf L},\phi,{\bf a}_1,{\bf b}_0, {\bf b}_1)$ is a positive smooth infinitely differential function of all its parameters. For a fixed $\Omega$, let us group these parameters in two vectors ${\bf x} = (\bf a_1, b_0,b_1)$ and ${\bf y} = (T_z, T_x, {\bf a_0}).$ We then formally define
\be
\begin{split}
    f({\bf x,y}) &= \mathcal{B}_\Omega({\bf L},\phi,{\bf a}_1,{\bf b}_0, {\bf b}_1)\\
    g({\bf y}) &=\max_{\bf x} f({\bf x,y}) =\beta_\text{max}({\bf L, a_0},\Omega).
\end{split}
\ee
We are interested in bounding $|dg({\bf y})|$ as a function of $d{\bf y}$.
Consider two values of the parameter, ${\bf y}_1$ and ${\bf y_2}$, and define
\be\begin{split}
    \bar{\bf x}_1 &= \text{argmax}_{\bf x} f({\bf x},{\bf y}_1) \\
    \bar{\bf x}_2 &= \text{argmax}_{\bf x} f({\bf x},{\bf y}_2), 
\end{split}
\ee
such that $g({\bf y}_1)= f(\bar{\bf x}_1,{\bf y}_1)$ and  $g({\bf y}_2)= f(\bar{\bf x}_2,{\bf y}_2)$. Without loss of generality we assume $g({\bf y}_1)\geq g({\bf y}_2)$ and consider the difference
\be\begin{split}
| g({\bf y_1}) - g({\bf y_2})| &= |f(\bar{ \bf x}_1, {\bf y_1} ) - f(\bar{ \bf x}_2, {\bf y_2} )|\\ &\leq
|f(\bar{ \bf x}_1, {\bf y_1} ) - f(\bar{ \bf x}_1, {\bf y_2} )|\\
&\leq \max_{{\bf x}} |f({ \bf x}, {\bf y_1} ) - f({ \bf x}, {\bf y_2} )|
    \end{split}
\ee
Taking the limit ${\bf y_2}\to {\bf y}_1$ we get 
\be
|dg\bf(y)|\leq \max_{\bf x} |\nabla_{\bf y} f({\bf x},{\bf y})\cdot d{\bf y} |.
\ee

Using the expression~\eqref{eq: Bell score full} for $\mathcal{B}_\Omega({\bf L},\phi,{\bf a}_1,{\bf b}_0, {\bf b}_1)$ we then get 
\be\label{eq: bound cont B}
|d\beta_\text{max}({\bf L},\phi)|\leq 
\left(\begin{array}{c}
1\\
1\\
\C_\Omega
\end{array}\right)
\cdot
\left(\begin{array}{c}
|dT_z|\\
|dT_x|\\
|d\phi|
\end{array}\right).
\ee

\subsection{Gradient of the goal function} 
\label{app: gradient of goal}
We will now combine all the elements to upper bound the gradient of the goal function
\be
G({\bf L}, \phi; \Omega, q) = H(\rho_E) - H(E|\hat{a}_0(q)) + t \beta_\text{max}({\bf L},\phi).
\ee
We parametrize the vector $\bf L$ with the help of the angles as in Eq.~\eqref{eq: L with angles}, such that the ``model'' is described by four angles
\be
({\bf L},\phi) \simeq {\bm \omega}= (\alpha,\mu,\xi, \phi).
\ee
We will bound the gradient of the goal function with respect to this parametrization.\\

We start with the gradient of the angle. It satisfies
\be
    dA =| A({\bf L}({\bm \omega}), {\bf L}({\bm \omega + {\bf n} d \omega}))|\leq   |\nabla A({\bm \omega})| d \omega  ,
\ee
with a unit vector ${\bf n}= (n_\alpha,n_\mu,n_\xi,0)$. For the fidelity one finds 
\be\begin{split}
   F({\bf L}({\bm \omega}), {\bf L}({\bm \omega + {\bf n} d \omega}))&= \sqrt{\bf L(\bm \omega)}\cdot \sqrt{\bf L(\bm \omega +d\bm \omega_i)}\\
    &\geq1 - \frac{3}{4} d\omega^2 +O(d\omega^3)
\end{split}
\ee
for $d\omega_i = d\alpha, d\mu$ or $d\xi$, one easily finds
\be
|\nabla A({\bm \omega})| \leq \sqrt{\frac{3}{2}}.
\ee
We thus get from Eq.~\eqref{eq: bound cont H}
\be\begin{split}
|d(H(\rho_E) - H(E|\hat{a}_0(q)))|\leq 10.023 \sqrt{\frac{3}{2}}\sqrt{d\alpha^2+d\mu^2 + d\xi^2} + 3 d\phi
\end{split}
\ee
and
\be\begin{split}
| \nabla (H(\rho_E) - H(E|\hat{a}_0(q)) |  \leq \sqrt{10.023^2 \frac{3}{2} + 9} < 12.7
\end{split}
\ee

For the gradient of the Bell score contribution $| \nabla \beta_\text{max}({\bf L},\phi)|$
we note that
\be
|dT_z|, |dT_x| \leq 2 \Big(|d\alpha| + |d\mu|+ |d\xi|\Big),
\ee
which implies form \eqref{eq: bound cont B} that
\be
| \nabla \beta_\text{max}({\bf L},\phi)| \leq t \sqrt{3 \times 4^2 +\C_\Omega^2} \leq 7 t
\ee
Finally, for the goal function 
\be\boxed{
|\nabla G| \leq  12.7 + 7\,  t }.
\ee

\subsection{Lipschitz function certified maximization on a compact space}
\label{app: algo}

\paragraph{General approach --} In this section section we explain how we obtain a certified maximum of a Lipschitz function $f:\mathbb{R}^n \to \mathbb{R}$ with a Lipschitz constant $\Lambda$, i.e. a bounded gradient $|\nabla f| \leq \Lambda$, on a closed domain
\begin{equation}
    \cD = \{{ x}=\colvec{x_1 \\ \vdots \\ x_n} \in\mathbb{R}^n\; |\; x_i \in [\tau_i,\mu_i],\;\forall\,i \}.
\end{equation}\\
We start by meshing $\cD$ into an hypercube grid graph $G(s)$ with element of size $s$. If possible, we take $s$ so that $s$ evenly divides $\mu_i-\tau_i$ for all $i$. We label the center of each hypercube $\vec{c}$ so that each first neighbor in a given direction is separated by $s$, e.g. for $\vec{c}$ and its neighbor $\vec{c'}$ in the direction $+\vec{e_1}$
\begin{equation}
    \vec{c'} = \colvec{c_1' \\ c_2' \\ \vdots \\ c_n'} = \vec{c}+s.\vec{e_1} = \colvec{c_1+s \\ c_2' \\ \vdots \\c_n'}.
\end{equation}
A hypercube of center $\vec{c}$, $h(\vec{c})$, element of G, is thus defined as
\begin{equation}
    h(\vec{c}) = \{x \in \mathbb{R}^n | c_i-\frac{s}{2}\leq x_i \leq c_i+\frac{s}{2},\,\forall i\in \{1,\ldots,n\}\}.
\end{equation} \\

Given the Lipschitz constant $\Lambda$, the function $f$ in a given hypercube $h(\vec{c})$ is upper bounded by
\begin{equation}
    f(x) \leq f(\vec{c})+\frac{\sqrt{n}\Lambda}{2}s, \quad \forall x\in h(\vec{c}).
\end{equation}
Trivially, an upper bound on the maximum of $f$ over $\cD$ is given by,
\begin{equation}
    \max_{x\in \cD} f(x) \leq \max_{\vec{c}\in G(s)} f(\vec{c}) + \frac{\sqrt{n}\Lambda}{2}s.
\end{equation}
The maximum of $f$ over $\cD$ can thus be obtained by taking the smallest possible step, i.e.
\begin{equation}
    \label{eq:optimal_max}
    \max_{x\in \cD} f(x) = \lim_{s\to 0}\left(\max_{\vec{c}\in G(s)}\left(f(\vec{c}) + \frac{\sqrt{n}\Lambda}{2}s\right)\right).
\end{equation}\\

\paragraph{Numerical realisation --} While \eq{optimal_max} gives the optimal maximum on $f$, it is however impracticable using numerical resources, due to the obvious need of discretization. A first naive approach is to relax the problem by setting a lower bound on $s$. This is, with the lower bound $s\geq\varepsilon$,
\begin{equation}
    \label{eq:naive_max}
    \max_{x\in \cD} f(x) \leq \lim_{s\to \varepsilon}\left(\max_{\vec{c}\in G(s)}\left(f(\vec{c}) + \frac{\sqrt{n}\Lambda}{2}s\right)\right).
\end{equation}
Resource wise, this method is cumbersome. Indeed, setting $\varepsilon$ to a small value compared to the domain space, i.e. $\varepsilon \ll \min_i(\tau_i-\mu_i)$, will result in a high number of hypercubes to explore. Furthermore, for a given step $s$, the number of hypercubes scales exponentially with the dimension $n$. \\

\paragraph{Speed-up using a guess on the maximum --} Providing a guess on the maximum of $f$ may significantly speed up the previous method. Such a maximum can be found using an optimization algorithm without guarantees of optimality (BFGS, CMA, and others...). Denote such a maximum as $\nu$.\\
We start by setting a not-too-small step, $s_0$, so that the number of hypercubes composing the grid graph, $G_0(s_0)$, is reasonnable. For each hypercube $h_0(\vec{c})$ we start by computing the potential maximum,
\begin{equation}
    \label{eq:max_hypercube}
    \xi^0(\vec{c}) = f(\vec{c})+\frac{\sqrt{n}\Lambda}{2}s_0.
\end{equation}
We then compare this value to $\nu$. In the case where $\nu>\xi^0(\vec{c})$, we pass to the next hypercube since the guessed maximum is higher than the potential maximum value of $f$ in $h_0(\vec{c})$. Otherwise, we mesh $h_0(\vec{c})$ into a new hypercube grid graph $G_1(s_1)$ of element of size $s_1=s_0/2$. For all of the new generated hypercubes, $h_1(\vec{c})\in G_1(s_1)$, we compute the potential maximum $\xi^1(\vec{c})$ using \eq{max_hypercube}. This method is applied recursively until either all new hypercubes of graph $G^k_m$ satisfy $\nu>h_m(\vec{c})$, or we reach the minimum step $s_m\leq\varepsilon$. An upper bound of $f$ is thus given by the maximum $\xi^m(\vec{c})$
\begin{equation}
    \max(\xi^m(\vec{c})) = f(\vec{c})+\frac{\sqrt{n}\Lambda}{2}s_m
\end{equation}
where $s_m=s_0/2^m$.
\newpage
\end{widetext}
\end{document}